\documentclass[a4paper,fleqn,usenatbib]{mn2e}

\usepackage{aas_macros}
\usepackage{color}
\usepackage{bm}
\usepackage{hyperref}
\usepackage{newtxtext,newtxmath}
\usepackage[T1]{fontenc}
\usepackage{times}
\usepackage{amssymb}
\usepackage{float}
\usepackage{graphicx}
\usepackage{epstopdf}
\usepackage{lscape}
\usepackage{threeparttable}

\def\be{\begin{equation}}
\def\ee{\end{equation}} 
\def\ltsima{$\; \buildrel < \over \sim \;$}
\def\lsim{\lower.5ex\hbox{\ltsima}}
\def\gtsima{$\; \buildrel > \over \sim \;$}
\def\gsim{\lower.5ex\hbox{\gtsima}}

\def\kms{km s$^{-1}$}

\title[HVS candidates in {\it Gaia} DR1/TGAS]{An artificial neural network to discover Hypervelocity stars: Candidates in {\it Gaia} DR1/TGAS} 

\author[Marchetti et al.]{T. Marchetti$^{1}$\thanks{E-mail: marchetti@strw.leidenuniv.nl}, E. M. Rossi$^{1}$, G. Kordopatis$^2$, A. G. A. Brown$^1$,  A. Rimoldi$^1$, \newauthor E. Starkenburg$^3$, K. Youakim$^3$ and R. Ashley$^4$ \\
$^1$Leiden Observatory, Leiden University, PO Box 9513 2300 RA Leiden, the Netherlands \\
$^2$Universit\'e C\^{o}te d'Azur, Observatoire de la C\^{o}te d'Azur, CNRS, Laboratoire Lagrange, France\\
$^3$Leibniz-Institut fur Astrophysik Potsdam (AIP), An der Sternwarte 16, 14482 Potsdam, Germany \\
$^4$Department of Physics, University of Warwick, Gibbet Hill Road, Coventry, CV4 7AL, UK \\
}

\begin{document}

\pagerange{\pageref{firstpage}--\pageref{lastpage}} \pubyear{2016}

\maketitle

\label{firstpage}



\begin{abstract} 
The paucity of hypervelocity stars (HVSs) known to date has severely hampered their potential to 
investigate the stellar population of the Galactic Centre and the Galactic Potential. The first {\it Gaia} data release (DR1, 2016 September 14) gives an opportunity to increase the current sample. The challenge is the disparity between the expected number of hypervelocity stars and that of bound background stars. 
We have applied a novel data mining algorithm based on machine learning techniques, an artificial neural network, to the Tycho-{\it Gaia} astrometric solution (TGAS) catalogue. With no pre-selection of data, we could exclude immediately $\sim 99\%$ of the stars in the catalogue and find 80 candidates with more than 90\% predicted probability to be HVSs, based only on their position, proper motions, and parallax. We have cross-checked our findings with other spectroscopic surveys, determining radial velocities for 30 and spectroscopic distances for 5 candidates. In addition, follow-up observations have been carried out at the Isaac Newton Telescope for 22 stars, for which we obtained radial velocities and distance estimates. 
We discover 14 stars with a total velocity in the Galactic rest frame $> 400$ \kms, and $5$ of these have a probability $>50\%$ of being unbound from the Milky Way. Tracing back their orbits in different Galactic potential models we find one possible unbound HVS with $v \sim 520$ \kms, $5$ bound HVSs, and, notably, $5$ runaway stars with median velocity between $400$ and $780$ \kms.
At the moment, uncertainties in the distance estimates and ages are too large to confirm the nature of our candidates by narrowing down their ejection location, and we wait for future {\it Gaia} releases to validate the quality of our sample. This test successfully demonstrates the feasibility of our new data mining routine.
\end{abstract}


\section{Introduction}

Observationally, hypervelocity stars (HVSs) are stars that can reach radial velocities in excess 
of the Galactic escape speed at their location, and whose trajectories are consistent with a 
Galactic Centre (GC) origin \citep{brown+05}. Currently, about $\sim 20$ unbound stars have been discovered \citep{brown+14}: most of them are late B-type stars ($\sim 2.5-4$ M$_{\odot}$) detected in the outer halo \citep[but
note][]{zheng+14} with velocities between  $\sim 300-700$ \kms \cite[see][for a review]{brown15}. These stars are in principle unique tools to gather information on the Galactic Centre stellar population and dynamics \cite[][e.g.]{madigan+14,zhang+13} and on the Galactic potential \cite[e.g.][]{gnedin+05,yu&madau07,perets+09}.
Using current data, a first proof of principle of how to get joint constraints on both environments was published in \citet{rossi+16}, and attempts to constrain the dark matter halo alone were performed by \citet{sesana+07} and \citet{fragione&loeb16}\footnote{See also \citet{gnedin+10}, who uses the velocity dispersion of halo stars from the hypervelocity star survey.}. These analyses however are severely hampered by the quality and quantity of the current small and rather biased sample. 

So far the most successful observational strategy has been to spectroscopically select late B-type stars in the outer halo. Since the stellar halo is dominated by an old stellar population, young stars likely come from other star-forming regions in the Galaxy, and a late B-type star has a long enough life-time ($\sim100-300$ Myr) to be able to travel to the outer halo from the Galactic Centre if its velocity is hundreds \kms. Most of the confirmed unbound HVSs have only radial velocity measurements and uncertainties in their photometric distances are large. Proper motions have been acquired with the Hubble Space Telescope for $16$ high velocity stars \citep{brown+15}, but even if the GC origin was confirmed for $13$ of these objects, uncertainties are still too large to precisely constrain their origin, and therefore to identify them as HVSs.

Recent years have seen an increasing effort to identify low mass HVSs in the inner Galactic halo. These searches use high proper motion or high radial velocity criteria, as it is not possible to spectroscopically single out these low mass stars in the halo, as is done for B-type HVSs. A few tens of candidates have been reported, but the large majority are bound and/or consistent with Galactic disc origin \cite[e.g.][]{li+12,palladino+14,ziegerer+15,vickers+15,hawkins+15,zhang+16, ziegerer+17}. Positive identification is prevented by large distance and proper motion uncertainties.

Major observational advancements in the field are therefore expected from the data taken by the ESA mission {\it Gaia}, launched on the 19th of December 2013 \citep{gaiaa, gaiab}. {\it Gaia} will attain an unparalleled astrometric measurement precision for a total of $\sim10^9$ stars in the Galaxy. In the end-of-the-mission data release, we anticipate a few hundred (a few thousand) HVSs within ~10 kpc from us, in the mass range $\sim1-10$ M$_\odot$, with relative error on total proper motion $<1\%$ ($< 10\%$), and that radial velocities will be measured for a subsample of these (Marchetti et al. in preparation). For brighter HVSs, accurate {\it Gaia} parallaxes can eliminate the large distance uncertainties in the existing sample, and for fainter stars calibrated photometric distances may eventually be used. 

The first data release (DR1) happened on September 14, 2016, and it contains the five-parameter astrometric solution (positions, parallaxes, and proper motions) for a subset of $\sim 2 \times 10^{6}$ stars in common between the Tycho-2 Catalogue and {\it Gaia} \cite[TGAS catalogue,][]{michalik+15, lindegren+16}. Radial velocity information is notably missing. Our expectation is that between $0.1-$ and a few unbound HVSs may be expected to be present in the catalogue, depending on the unknown mass distribution and star formation history in the Galactic Centre (Marchetti et al. in preparation).

In this paper, we report a systematic search for HVSs in DR1. We use an artificial neural network (\S \ref{sec:n_network}), which is first applied to the TGAS subset of the {\it Gaia} catalogue without any prior constraints placed on stellar properties to select HVS candidates (\S \ref{sec:DR1}). We then cross check our sample of best candidates with published spectral catalogues to acquire radial velocity and spectroscopic distance information (\S \ref{sec:RV}). We further proceed to describe the radial velocity follow-up observations for candidates with no published radial velocity and observable by the Isaac Newton Telescope (INT) (\S  \ref{sec:INT}). In \S \ref{sec:dist} we describe our Bayesian approach to determine distances, and then in \S \ref{sec:results} we present our results for HVS candidates in terms of total velocity and ejection location. We sort and characterize candidates in \S \ref{sec:candidates}, and discuss their implications in \S \ref{sec:discussion}.


\section{Data Mining algorithm}
\label{sec:n_network}

Hypervelocity stars are rare objects, that occur in the Galaxy at an uncertain rate roughly between $10^{-5}-10^{-4}$ yr$^{-1}$ \citep{hills88, perets+07, zhang+13, brown+14}. Considering the magnitude limit of {\it Gaia} and different assumptions on the population of binaries in the GC, such a rate implies only $\sim 0.1-1$ HVSs for every $10^{6}$ stars in the final {\it Gaia} catalogue (Marchetti et al. in preparation). In particular for the TGAS catalogue, we expect to find at most a few HVSs (Marchetti et al. in preparation), although a larger number of slower stars generated via the same mechanism (called {\it``bound HVSs"}) are also expected \citep{bromley+06, kenyon+08}. Thus, {\it Gaia} can deliver a HVS sample that represents a huge leap in data quality and quantity, but building it requires careful data mining, especially since radial velocity measurements are currently missing. 

The TGAS subset of {\it Gaia} DR1 provides the five-parameter astrometric solution for roughly two million objects, therefore we choose to build a data mining routine based only on the astrometric properties of the stars: position on the sky $(\alpha, \delta)$, parallax $\varpi$, and proper motions $\mu_{\alpha *}, \mu_\delta$. This approach allows us to not make any a priori assumption on the stellar nature of HVSs, avoiding photometric and metallicity cuts which might bias our search towards particular spectral types, and lead to a sample which may not reflect the properties of the binary population in the Galactic Centre. Recent studies have shown indeed how the GC is a complex environment in which different stellar populations coexist and interact, and many properties (mass function, metallicity, binarity) are missing or poorly constrained due to observational limitations (see \cite{genzel+10} for an exhaustive review). The nuclear star cluster surrounding the central massive black hole has also undergone several star formation episodes throughout its lifetime, which might have changed and influenced the stellar population and mass function \citep{genzel+10, pfuhl+11}.

We have therefore chosen to build a data mining routine based on a machine learning algorithm, an {\it artificial neural network}. Our chosen approach is a \emph{supervised learning} problem: we present the algorithm with examples and their desired output (\emph{training set}), and we let the algorithm learn the best function mapping inputs into outputs. We decided for a binary classification problem: the desired output of the algorithm is $0$ for a ``normal" background star, and $1$ for a HVS. When we apply the classification rule to a new unlabelled example we can then interpret its output as the probability of that star being a HVS \citep{saerens+02}.

We now start introducing neural networks, with a brief summary on the main idea behind this algorithm. Next in \S \ref{sec:train} we discuss how we build our training set, and finally in \S \ref{sec:opt} and \S \ref{sec:perf} how we optimize and determine the performance of the network based on the results on subsets of the data which were not used for the training.

\subsection{Artificial Neural Networks}

Artificial neural networks have been largely used in different branches of science for their ability to provide highly non-linear mapping functions, and for their intrinsic capacity to generalize: to provide reasonable outputs for examples not encountered while training the algorithm (see \cite{haykin} for an exhaustive explanation of neural networks). This latter property is particularly important for our goal, since our training set consists of mock data (see \S \ref{sec:train}), and therefore we want to be flexible enough to find HVSs even if the real population is not perfectly represented by our simulations, which necessarily rely on several hypotheses and assumptions (see \S \ref{sec:train}).

We have developed from scratch an artificial neural network with five input units (the astrometric parameters), two hidden layers of neurons, and a single output neuron for binary classification. Each neuron of the network is a computational unit which outputs a non-linear function\footnote{In the following, we will use superscripts in round brackets to refer to a particular vector, and subscripts to specify its components.} $f(v)$, where $v$ is a linear combination of the $j$-th input M-dimensional vector $\bm{x}^{(j)}$ with some weight vector $\bm{\omega}$:
\begin{equation}
\label{nn:o}
v_j(\bm{x}^{(j)};\bm{\omega}) = x_0\omega_0 + \sum_{i=1}^\mathrm{M} x^{(j)}_i \omega_i ,
\end{equation}
where $x_0 \equiv 1$ is referred to as the \emph{bias unit}. In analogy with the brain architecture, the components $\omega_i$ are usually referred to as {\it synaptic weights}. A typical choice for $f$ is a sigmoid function. We choose:
\begin{equation}
\label{nn:hyp}
f(v) = a \tanh (b v) ,
\end{equation}
with $a = 1.7159$ and $b = 2/3$. This activation function outputs real numbers in the interval $[-a, a]$, and satisfies several useful properties: it is an odd function of its argument; $f(1)=1$ and $f(-1)=-1$; its slope at the origin is close to unity; and its second derivative attains its maximum value at $x = 1$. This choice has been shown to yield faster convergence than the usual logistic function, avoiding driving the hidden neurons into saturation \citep{LeCun}.

For neurons in the first hidden layer the input $\bm{x}^{(j)}$ is just the data vector containing the $\mathrm{M} = 5$ astrometric parameters for the $j$-th training example: $\bm{x}^{(j)} = ( \alpha_j, \delta_j, \varpi_j, \mu_{\alpha* j}, \mu_{\delta j} )$, therefore the summation in Equation \ref{nn:o} extends over $i = 1, \dots, 5$. For neurons in the second layer the input $\bm{x}^{(j)}$ is the M$_1$-dimensional vector output by the first layer of M$_1$ neurons, and the summation extends to $\mathrm{M} = \mathrm{M}_1$. Finally, the single neuron in the output layer takes in input a M$_2$-dimensional vector, with M$_2$ equal to the number of neurons in the second hidden layer, and in summation $\mathrm{M} = \mathrm{M}_2$. We call $D_j(\bm{\omega}) \in \mathbb{R}$ the final output of the neural network for the $j$-th example. 

The training process consists in finding the vector of synaptic weights $\bm{\omega}$ which minimizes the total cost function
\begin{equation}
\label{eq:J}
J(\bm{\omega}) \propto \sum_{j=1}^N (D_j(\bm{\omega}) - y_j)^2,
\end{equation}
which is just the sum over all the $N$ examples of the squared difference between the output of the neural network $D_j(\bm{\omega})$ and the desired output $y_j$ of the labelled training example. The value of each synaptic weight is initialized with a random number drawn from a uniform distribution in the interval $[-\sigma_\omega, \sigma_\omega]$, with $\sigma_\omega = m_*^{-1/2}$, where $m_*$ is the number of connections feeding into the corresponding layer of neurons \citep{LeCun2012}. The weights optimization is then performed with an adaptive stochastic (online) gradient descent method, using a specific learning rate $\eta_k$ for each synaptic weight: the AdaGrad implementation \citep{duchi}. We use the following iterative rule for the $t$-th update of the $k$-th weight $\omega_k$ \citep{singh}:
\begin{equation}
\label{eq:adagrad}
\Delta \omega_k(t) = -\eta_k(t)g_k(t) =  -\frac{\eta_0}{\sqrt{\sum_{i=1}^t(g_k(i))^2}}g_k(t),
\end{equation}
where $\eta_0 > 0$ is called the {\it global learning rate}, $\bm{g}$ is the gradient of the cost function in Equation \ref{eq:J} (derivatives with respect to the weight vector $\bm{\omega}$), and the denominator is the norm of all the gradients of the previous iterations. The adopted value for $\eta_0$ is discussed in \S \ref{sec:opt}, while the gradient of the cost function is estimated with a back-propagation algorithm (see \cite{LeCun2012} for tips on an efficient implementation, essential when dealing with large datasets). 

\subsection{Building the Training Set}
\label{sec:train}

We train the artificial neural network on a simulated end-of-mission {\it Gaia} catalogue for the Galaxy: the {\it Gaia} Universe Model Snapshot \citep[GUMS,][]{robin+12}, where we inject {\it mock} HVS data with errors on all astrometric and photometric measurements. A detailed description of how we construct our mock HVS will be the focus of an upcoming paper, and here we only briefly summarize our procedure. In the following we will adopt the Hills mechanism for modelling our mock population of HVSs, involving the disruption of a binary star by the Massive Black Hole (MBH) at the centre of our Galaxy \citep{hills88}. 

We explore the space $(l, b, d, M)$ to populate each position in Galactic coordinates on the sky $(l,b)$ with stars in a mass range $M \in [0.1 - 9]$ $M_\odot$ and in a distance range $d \in [0, 40]$ kpc from us. We adopt a step of $\sim 9^\circ$ in Galactic longitude $l$, $\sim 4.5^\circ$ in Galactic latitude $b$, $\sim 1$ kpc in distance $r$, and $\sim 0.2$ M$_\odot$ in mass. We draw velocities from an ejection velocity distribution  which analytically depends on the properties of the original binary approaching the massive black hole \citep{sari+10, kobayashi+12, rossi+14}\footnote{Rigorously, there should be a numerical factor in front of Equation \ref{eq:vej}, depending on the detailed geometry of the three-body encounter. This factor has been shown to be $\sim 1$ when averaged over the binary's phase \citep{rossi+14}.}: 
\begin{equation}
\label{eq:vej}
v_\mathrm{ej} = \sqrt{\frac{2 G m_c}{a}} \Biggl(\frac{M_\bullet}{m_T} \Biggr)^{\frac{1}{6}},
\end{equation}
where $m_c$ is the mass of the star that remains bound to the MBH after the binary is disrupted, $m_T = M + m_c$ is the total mass of the disrupted binary, and $M_\bullet = 4.0 \times 10^6$ M$_\odot$ is the mass of the MBH in our Galaxy \citep{ghez+08, gillessen+09, meyer+12}. Following \cite{rossi+14, rossi+16}, we model binary distributions for semi-major axis $a$ and mass ratio $q$ as power-laws: $f_a \propto a^\alpha$, $f_q \propto q^\gamma$, with exponents $\alpha = -1$ \cite[\"{O}pik's law,][]{opik} and $\gamma = -3.5$. This combination has been shown to result in a good fit between the observed sample of late B type HVSs in \cite{brown+14} and the prediction of the Hills mechanism for reasonable choices of Milky Way potentials \citep{rossi+16}. The total velocity $v$ of the HVS is then computed decelerating the star in a given Galactic potential (refer to \S \ref{sec:traceback}, Equations \ref{eq:bulge}-\ref{eq:nfw} for details on the adopted fiducial Milky Way potential).

For each star we compute the combination of proper motions and radial velocity which are consistent with an object moving radially away from the Galactic Centre, and we correct those values for the motion of the Sun and of the local standard of rest (LSR) \citep{schonrich}. We then roughly estimate the flight time from the GC to the given position in Galactocentric coordinates $r_\mathrm{GC}$ as $t_F = r_\mathrm{GC}/v_F$, where $v_F$ is an effective velocity equal to the arithmetic mean between the ejection velocity and the decelerated velocity at the star's position. The age of the star is then computed summing the flight time and the age of the star at its ejection. The latter is computed as a random fraction of its main sequence (MS) lifetime \citep{brown+14}, and the time spent on the MS is computed using analytic formulae in \cite{hurley+00}. We assume a super-solar metallicity $[M/H] = 0.4$, which corresponds to the mean value of the distribution in the GC \citep{do+15}. Each star is evolved up to its age using the fast parametric stellar evolution code SeBa \citep{portegieszwart+96, portegieszwart+09} to obtain its radius, effective temperature, and mass, which we use to identify the best-matched stellar spectrum from the BaSeL 3.1 stellar spectral energy distribution (SED) libraries \citep{westera+03} via chi-squared minimization. For each position of the sky we assess dust extinction using a three-dimensional Galactic dust model \citep{drimmel+03}, and integrating the reddened flux in the respective passbands we estimate the magnitudes in the {\it Gaia} $G$ band and in the Johnson-Cousins $V$, $I_c$ bands. We finally use the python toolkit PyGaia\footnote{\url{https://github.com/agabrown/PyGaia}} to estimate the errors on the astrometry with which {\it Gaia} would observe these objects. The errors are functions of the magnitude of the star, its color index $V-I_c$, and the ecliptic latitude $\beta$, the latter determining the number of observations of the object according to the satellite's scanning strategy. 

Parallax and proper motions of each source are then replaced by drawing a random number from a Gaussian distribution centred on the nominal value and with standard deviation equal to the estimated uncertainty. This approach has two main advantages: it allows us to obtain negative parallaxes (which are present in the real {\it Gaia} catalogue) for faint objects with non-negligible relative errors on parallax; and it helps us mitigate the effect of the spatial grid in distance used for generating mock stars, preventing the algorithm from driving the learning rule towards discrete, fixed values in parallax.

We can therefore build a mock catalogue of HVSs, which we use for the training of the artificial neural network. We combine mock positions, parallaxes and proper motions of HVSs and ``normal" background stars randomly picked from the GUMS in a single stellar catalogue, consisting of a total of $\sim 2.5 \times 10^6$ objects ($\sim 25\%$ HVSs, label = $1$; $\sim 75\%$ {\it Gaia} stars, label = $0$). We randomly split stars of the catalogue into a {\it training} set ($\sim 60\%$ of the catalogue), a {\it cross-validation} set ($\sim 20\%$ of the catalogue), and a {\it test} set ($\sim 20\%$ of the catalogue). The training set consists of the examples the algorithm will learn from, the cross-validation set is used to optimize hyperparameters (see \S \ref{sec:opt}), while we use the test set to determine the performance of the neural network (see \S \ref{sec:perf}). The use of different examples for performing these tasks is extremely useful to prevent overfitting and to ensure generalization. All features (five parameters) of the complete catalogue have been scaled in such a way to have mean of $0$ and variance of $1$, to achieve a faster convergence of the stochastic gradient descent algorithm \citep{LeCun2012}.

\subsection{Optimization of the Algorithm}
\label{sec:opt}

The effectiveness of a neural network, as the majority of machine learning algorithms, critically depends on the choice of the so-called \emph{hyperparameters}, several parameters that need to be carefully tuned in order to achieve the best compromise between the algorithm performance, the time needed for its training, and its ability to generalize to new input data. We identify three hyperparameters in our algorithm: the number of neurons in the first hidden layer M$_1$, the number of neurons in the second hidden layer M$_2$, and the global learning rate $\eta_0$ for the adaptive stochastic gradient descent (see Equation \ref{eq:adagrad}).

A systematic grid search in the hyperparameter space to determine the best combination is not feasible because of time limitations and computational power. We use the pyswarm\footnote{\url{https://github.com/tisimst/pyswarm/}} implementation of a Particle Swarm Optimization (PSO) algorithm \citep{pso} to explore the space (M$_1$, M$_2$, $\eta_0$) with $20$ test particles. The algorithm iteratively adjusts particles' positions towards the minimum value attained by the cost function, with a velocity proportional to the distance from this extremum. Since each iteration involves the full training of the algorithm in order to determine the value of the cost function, we choose to apply PSO to a limited sample of the training set ($1000$ random training examples), and then we select the combination of parameters which results in the best performance on the full cross-validation set, defined in terms of the Matthews correlation coefficient MCC \citep[][see next subsection]{Matthews}. The PSO algorithm converges to the following values: M$_1 = 119$, M$_2 = 95$, $\eta_0=0.071$\footnote{We initially included the \emph{regularization parameter} $\lambda$ as a $4$th hyperparameter, but due to time limitation with the PSO we decided to discard it, since several tests showed that it always converged to values close to zero. A value $\lambda \sim 0$ is an indication that the algorithm is not overfitting the training set.}.

\subsection{Performance of the Algorithm}
\label{sec:perf}

As mentioned before, we choose a stochastic gradient descent optimization to minimize the global cost function. Because of the intrinsic randomness of this algorithm, we train the neural network several times on the complete training set, shuffling the order of the presented example units during each training. Plotting learning curves (the value of the cost function versus the number of training examples presented to the network), we find that $8$ complete iterations are enough to reach a minimum in both the training and cross-validation cost functions, again confirming that overfitting is not an issue.

We determine the performance of the algorithm on the test set by computing two different error metrics: the \emph{Matthews correlation coefficient} MCC \citep{Matthews} and the \emph{F$_1$ score}. Calling TP and TN (FP and FN) respectively the number of true (false) positives and negatives of the confusion matrix on the test set, error metrics are computed as:
\begin{equation}
\label{eq:F1}
\mathrm{F}_1 \equiv 2 \frac{\mathrm{P} \mathrm{R}}{\mathrm{P}+\mathrm{R}},
\end{equation}
\begin{equation}
\label{eq:MCC}
\mathrm{MCC} \equiv \frac{\mathrm{TP} \ \mathrm{TN} - \mathrm{FP} \ \mathrm{FN}}{\sqrt{(\mathrm{TP} + \mathrm{FP})(\mathrm{TP} + \mathrm{FN})(\mathrm{TN} + \mathrm{FP})(\mathrm{TN} + \mathrm{FN})}},
\end{equation}
where P and R are called, respectively, \emph{precision} and \emph{recall}, and they are defined as $\mathrm{P} \equiv \mathrm{TP}/(\mathrm{TP}+\mathrm{FP})$, $\mathrm{R} \equiv \mathrm{TP}/(\mathrm{TP}+\mathrm{FN})$. The F$_1$ score assumes values in $[0,1]$ while the MCC in $[-1,1]$, and in both cases a value of $1$ corresponds to a perfect classifier (diagonal confusion matrix). At the end of the training, we obtain the following values on the test set: F$_1 \sim \mathrm{MCC} \simeq 0.95$.

\begin{figure}
\centering
\includegraphics[width=0.5\textwidth]{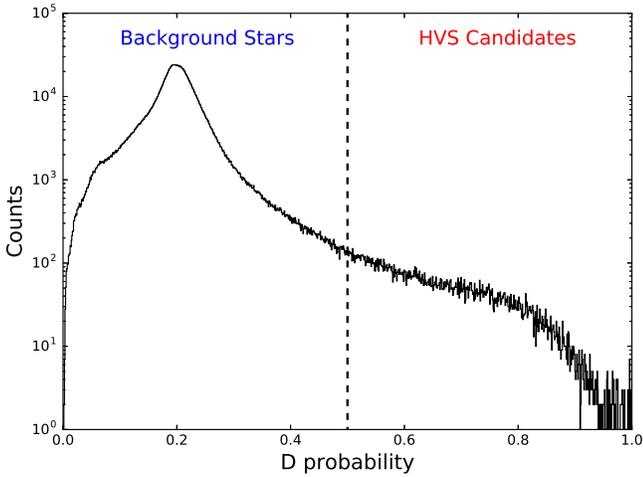}
\caption{Histogram of the probability $D$ of an object of being a HVS (output of the neural network), for all $\sim 2$ million stars in the TGAS subset of {\it Gaia} DR1. A dashed vertical line marks the decision boundary $D = 0.5$.}
\label{fig:NNprob}
\end{figure}

\section{Application to {\it Gaia} DR1}
\label{sec:DR1}

Once we have fully trained the neural network on the training set, determining the optimal values for the synaptic weights, we apply the classification rule to real unlabelled data to search for HVS candidates.  The application of the neural network to the full TGAS subset of {\it Gaia} DR1 ($2 057 050$ sources) results in $22 263$ stars with a predicted probability $>50\%$ of being a HVS, $\sim 1\%$ of the original dataset. The histogram of the output probability $D$ given by the neural network on the full TGAS catalogue is shown in Figure \ref{fig:NNprob}. To further reduce the sample of HVS candidates and to have reliable distance determinations, we filter out stars with a relative error on parallax $|\sigma_\varpi/\varpi| > 1$, obtaining a total of $8 175$ objects ($\sim 0.4\%$ of the original catalogue).

In these first cuts no information on the measured uncertainties is used to determine the probability of a star being a HVS. We subsequently include errors with a Monte Carlo (MC) simulation, randomly drawing one thousand realizations of the astrometry (parallax and proper motions) of each star from a Gaussian distribution centred on the nominal mean value and with a standard deviation equal to the corresponding quoted random uncertainty. This allows us to get for each star in TGAS a probability distribution of the output $D$ of the neural network, which can then be characterized by its mean $\bar{D}$ and standard deviation $\sigma_D$. As a final cut, we select only stars with $\bar{D}-\sigma_D>0.9$, for a total of $80$ best HVS candidates, $\sim 0.004\%$ of the original catalogue size.

We stress that all our cuts rely on the astrometry of the objects, without any prior assumption on the spectral type, photometry or more in general stellar properties of the selected best sample, and without any information on radial velocities.

\section{Acquiring spectral information}
\label{sec:RV}
To confirm or reject a candidate in our quest for HVSs, a measure of the star {\it total} velocity is necessary. In the following, we will describe how we obtained reliable heliocentric radial velocities (HRVs) for 47 stars out of the 80 candidates.

\subsection{Catalogue cross-matching}
\label{sec:X_match}
Our final sample has been cross-matched with several spectroscopic surveys of the Milky Way, covering both the Northern and Southern hemisphere\footnote{RAVE DR4 and DR5 \citep{kordopatis+13, kunder+16}, {\it Gaia}-ESO DR2 \citep{gilmore+12, randich+13}, LAMOST DR1 and DR2 \citep{lamostdr1}, GALAH \citep{galah}, APOGEE DR13 \citep{zasowski+13}.}. We find a total of 30 stars in common: a subsample of these (5 stars) have both radial velocity and spectroscopic distance from the RAdial Velocity Experiment (RAVE) DR4 and/or DR5 \citep{kordopatis+13, kunder+16}. 

\subsection{Follow-up observations with the INT}
\label{sec:INT}

We successfully applied for director's discretionary time at the Isaac Newton Telescope (INT) in La Palma, Canary Islands, where we followed up spectroscopically $22$ HVS candidates on the night of the $5$th of October, 2016. We used the Intermediate Dispersion Spectrograph (IDS) with the RED+2 CCD, in combination with the R1200R grating, a 1.35'' slit width, and the GG495 sorting order filter. This set-up provided an effective spectral range of $\sim 8000 - 9150$ \AA \ and a resolution at 7000 \AA \ of 6731 over 2 pixels at the detector. We ensured that all observed spectra had a S/N of at least 50.

\subsubsection{Spectra reduction}

The spectra were reduced using the Image Reduction and Analysis Facility \citep[IRAF,][]{Tody86} software package. The reduction procedure included pre-processing (bias and flat field corrections), spectrum extraction, wavelength calibration, heliocentric radial velocity correction, and continuum normalisation.

\subsubsection{Radial velocities, atmospheric parameters and spectroscopic distance determination}
\label{sec:stel_params}

A first pass for radial velocity determination is performed by using the python routine pyasl.crosscorrRV, adopting a Solar template as reference, and errors in radial velocities are obtained following \cite{zucker03}. In order to obtain the effective temperature, surface gravity and metallicity of the stars, the same pipeline as the one used in RAVE \citep{kordopatis+11, kordopatis+13} has been applied to the spectra. This implies keeping only the wavelength range $\lambda \lambda = [8450.80-8746.55]\AA$, removing the cores of the Calcium triplet lines (to avoid a mismatch between the synthetic templates used by the pipeline, computed assuming Local Thermodynamical Equilibrium, and the cores of the lines formed in Non LTE), and convolving the observations to a resolution of $R=7500$.  The output of the pipeline is then calibrated  using the formulas presented in \cite{kunder+16}.

Our final radial velocities are obtained through the cross-correlation of a synthetic spectrum of the best-fit parameters to the observed spectrum. This cross-correlation is done with the package fxcor in IRAF \citep{Tody86}. Both the observed and synthesized spectrum are continuum normalized before cross-correlation and we use a Gaussian fit to all points with a correlation of 0.5 or higher to determine the radial velocity and its corresponding measurement uncertainty. During the observations a sample of 14 radial velocity standard stars from \cite{soubiran+13} were observed with the same setup and matched closely in sky position to our program targets to check the accuracy of our determined radial velocities. We find that there is a good agreement between the literature values and our radial velocities. A mean offset of $\sim 0.1$ \kms assures us that there are no significant systematic effects. However, the rms variance between the literature values and our radial velocity determinations of $2.7$ \kms is significantly larger than the median measurement uncertainty in the cross-correlation alone, which is only $1.1$ \kms. We thus adopt an uncertainty floor of $2.5$ \kms and add this in quadrature to our measurement uncertainties. Although we believe the radial velocities derived in this second iteration to be more precise than the first pass radial velocities due to the use of a synthetic spectrum that fits the stellar parameters, we note that the results presented in this paper are robust to the use of either set of radial velocities.

To obtain the spectroscopic distances of the stars, the calibrated stellar parameters are projected on Padova isochrones spanning ages from 100 Myr to 13.5 Gyr, with a step of 0.1 Gyr and a metallicity range between $-2.2$\,dex and $+0.2$\,dex. This allows us to obtain the absolute magnitudes in several photometric bands as in \citet{kordopatis+11b, kordopatis+13c, kordopatis+15}, and an estimation of the age of the stars as in \citet{kordopatis+16, magrini+17}. The distances are then obtained using the distance modulus in the $J$ band, and assuming $A_J = 0.709 \ E(B-V)$ \citep{schlafly+11}, where $E(B-V)$ are the Schlegel extinctions towards each line-of-sight.

Kinematic properties from {\it Gaia} TGAS, radial velocities and stellar parameters derived from spectra of observed HVS candidates are presented in Table \ref{TAB:int}. For a precise cross-match with future {\it Gaia} releases and other Milky Way surveys, in Appendix \ref{app:GaiaID} we report the {\it Gaia} and Hipparcos identifier of all the observed sources. We note that for 4 stars out of 22, the pipeline has not converged  (quality flag $F = 1$, see Table\,\ref{TAB:int}) and therefore are excluded from the following analysis. Furthermore, visual inspection of TYC 2292-1267-1  (quality flag $F = 3$), shows a clear mismatch between the observed spectrum and the fitted template, and therefore was discarded as well.

\begin{figure}
\centering
\includegraphics[width=0.5\textwidth]{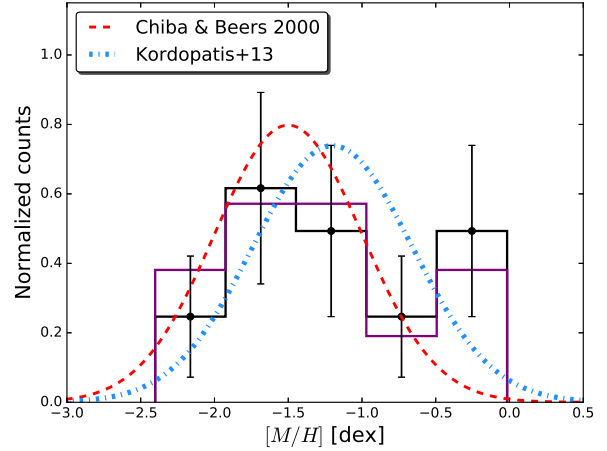}
\caption[]{Normalized $[M/H]$ distribution for the observed HVS candidates, with error bars computed assuming Poisson noise. For a visual comparison, we overplot with a red dashed (blue dot-dashed) line the inner stellar halo metallicity, modelled as Gaussian with mean and standard deviation from \cite{chiba&beers00} (\cite{kordopatis+13b}). Purple line shows the normalized  $[M/H]$ distribution for high-velocity candidates (see Table \ref{TAB:highV}).}
\label{fig:MH}
\end{figure}

\begin{figure}
\centering
\includegraphics[width=0.5\textwidth]{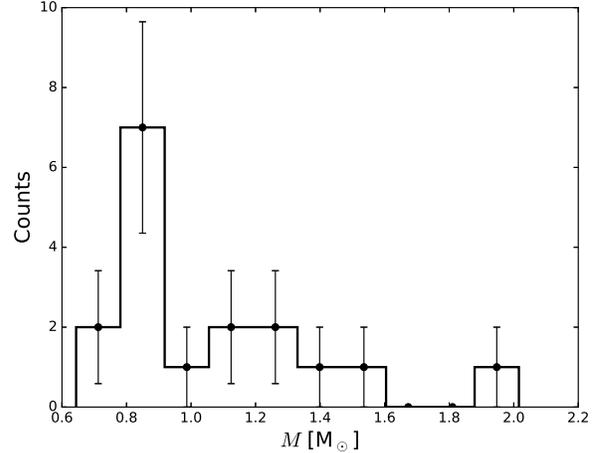}
\caption{Mass distribution for the observed HVS candidates, with error bars computed assuming Poisson noise. The peak of the distribution is $\sim 0.85$ M$_\odot$.}
\label{fig:Mass}
\end{figure}

The metallicity and mass distribution are shown, respectively, in Figure \ref{fig:MH} and \ref{fig:Mass}.  The mean metallicity of our sample is $-1.2$ dex, consistent with the inner Galactic halo distribution, dashed \citep{chiba&beers00} and dot-dashed \citep{kordopatis+13b} lines, but a total of 6 stars have $[M/H] > -0.5$ dex, and one candidate, TYC 3945-1023-1, has $[M/H] = -0.02 \pm 0.12$ dex. Most of the stars have masses slightly below the Solar value, with a peak of the distribution at $M \sim 0.85$ M$_\odot$, and a single star with $M \sim 2$ M$_\odot$: TYC 4032-1542-1. We can see that our sample is very different from the late B-type HVS candidates discovered in \cite{brown+14}. Considering the age estimates in Table \ref{TAB:int}, we note that the peak of the mass distribution is at the main-sequence turn-off of the stellar halo. Stars of this type have been used to trace the stellar halo because of their luminosity \citep[e.g.][]{cignoni+07}.

\section{Distance estimation}
\label{sec:dist}

Most of the stars in {\it Gaia} DR1 have non-negligible parallax errors. Therefore simply estimating distances as the inverse of parallax leads to biased results due to this highly non-linear transformation \citep{bailer-jones, astraatmadjaI}. Additionally it can not be applied to negative parallaxes, which are present in our sample. In order to correctly take into account correlations between astrometric parameters supplied by the {\it Gaia} catalogue (parameter correlations may have an important impact on our results since we are implementing Monte Carlo simulations), we choose not to use the distance catalogue presented in \cite{astraatmadjaII}, but to implement our own Bayesian approach, generalizing their method and considering covariances.

Assuming Gaussian noise for astrometric parameters, we model the likelihood for the triplet $\{\mu_{\alpha *}, \mu_\delta, \varpi \}$ as a multivariate normal distribution with mean vector:
\begin{equation}
\label{eq:meanvec}
\bar{x} = (\mu_{\alpha *}, \ \mu_\delta, \ 1/d),
\end{equation}
and with covariance matrix:
\begin{equation}
\label{eq:covmatr} 
\footnotesize
\arraycolsep=3pt
\thickmuskip =1.5mu
\Sigma = {}
\left(
\begin{array}{@{}ccc@{}}
\sigma_{\mu_{\alpha *}}^2 & \sigma_{\mu_{\alpha *}} \sigma_{\mu_\delta} \rho_{\mu_{\alpha *},\mu_\delta} & \sigma_{\mu_{\alpha *}} \sigma_\varpi \rho_{\mu_{\alpha *},\varpi} \\
\sigma_{\mu_{\alpha *}} \sigma_{\mu_\delta} \rho_{\mu_{\alpha *},\mu_\delta} & \sigma_{\mu_\delta}^2 & \sigma_{\mu_{\delta}} \sigma_\varpi \rho_{\mu_\delta,\varpi} \\
\sigma_{\mu_{\alpha *}} \sigma_\varpi \rho_{\mu_{\alpha *},\varpi} & \sigma_{\mu_\delta} \sigma_\varpi \rho_{\mu_\delta,\mu_\varpi} & \sigma_\varpi^2 
\end{array}
\right),
\end{equation}

\begin{landscape}
\begin{table}
\footnotesize
\centering
\caption{Kinematic and observational properties of 22 HVS candidates spectroscopically followed-up with the INT telescope.}
\label{TAB:int}

\begin{threeparttable}
\begin{tabular}{lcccccccccccl}
\hline

Tycho 2 ID & (RA, dec) & $\varpi$ & $\mu_{\alpha*}$ & $\mu_{\delta}$ &
HRV & T$_\mathrm{eff}$ & $\log g$ & $[M/H]$ & $d_\mathrm{spec}$ & $M$ & $t_\mathrm{age}$ & $F$ \\
 & (deg) & (mas) & (mas yr$^{-1}$) & (mas yr$^{-1}$) & (km s$^{-1}$) & (K) & (cm s$^{-2}$) & (dex) & (pc) & (M$_\odot$) & (Gyr) & \\

\hline

2282-208-1 & (16.81855, 33.66159) & $2.17 \pm 0.31$ & $202.643 \pm 1.213$ & $-62.458 \pm 0.398$&  $-0.61 \pm 1.29$ & $5936 \pm 136$ & $3.8 \pm 0.2$ & $-1.35 \pm 0.19$ & $606 \pm 152$    &  $0.92 \pm 0.17$ & $10.4 \pm 3.8 $ & 1 \\
2292-1267-1 & (20.86832, 31.78668) &$1.78 \pm 0.35$ &$90.782 \pm 0.969$ &$-15.275 \pm 0.644$& $158.93 \pm 5.99$&  $7861 \pm 83$& $4.0 \pm 0.2$& $-0.20 \pm 0.12$& $340 \pm 69$  &$1.70 \pm 0.14$ &$ 0.9 \pm 0.2 $ & 3 \\
2298-66-1 & (25.30039, 33.51859) & $2.45 \pm 0.34$&$178.060  \pm 1.213$ &$-19.060 \pm 0.319$& $-31.66 \pm 2.78$& $ 5925\pm 328$ & $3.8 \pm 0.5$ &$-2.08 \pm 0.26$& $754 \pm 569$ &$0.95 \pm 0.23$& $8.2 \pm 4.5 $ & 0\\
2320-470-1 & (31.29, 35.6289) &$2.06 \pm 0.27$ &$106.443 \pm 0.967$ & $6.138 \pm 0.290$&$-43.08 \pm 1.32$ &  $5730 \pm 214$ & $3.4 \pm 0.5$& $-3.29 \pm 0.27$& $1240 \pm 650$ &$1.00 \pm 0.21$&$6.9 \pm 4.0 $ & 1 \\
2376-691-1 & (66.43652, 33.59088)  &$1.17 \pm 0.29$ &$62.060 \pm 2.077$ &$-9.137 \pm 1.547$& $22.02 \pm 1.63$ & $5260 \pm 74$& $3.5 \pm 0.2$&$-0.67 \pm 0.11$& $249 \pm 64$  &$1.22 \pm 0.19$& $4.8 \pm 3.8 $ & 2\\
2393-1001-1\footnotemark& (78.45391, 32.03592) & $2.21 \pm 0.28$&$121.797 \pm 1.710$ &$-46.605 \pm 1.158$& $-106.50 \pm 0.94$ &  $4651 \pm 1.158$ & $0.6 \pm 0.2$&$-2.40 \pm 0.14$& $3036 \pm 462$ &$0.85 \pm 0.26$ &$7.6 \pm 2.2 $ & 0 \\
2818-556-1 & (23.79684, 40.43319)  &$2.56 \pm 0.37$ &$147.979 \pm 1.369$ &$-41.076 \pm 0.468$& $-92.17 \pm 1.42$ & $5734 \pm 63$ & $3.4 \pm 0.2$ &$-0.98 \pm 0.17$& $686 \pm 153$ & $1.30 \pm 0.18$&$3.4 \pm 2.9 $ & 2\\
2822-1194-1 & (23.14799, 42.03068) & $1.85 \pm 0.64$&$88.644 \pm 1.849$ &$2.063 \pm 0.496$&  $-23.19 \pm 1.87$& $6403 \pm 116$ & $4.2 \pm 0.2$ &$-0.48 \pm 0.12$& $532 \pm 160$ & $1.10 \pm 0.09$& $1.8 \pm 2.5 $ & 0\\
3163-1181-1 & (303.97045, 44.18376) &$2.30 \pm 0.25$ & $156.232 \pm 1.116$&$67.079 \pm 1.026$& $-194.08 \pm 1.61$ &  $5570 \pm 74$&$3.4 \pm 0.2$ & $-0.30 \pm 0.11$& $463 \pm 85$ &$1.59 \pm 0.17$ & $2.0 \pm 1.1 $ & 1\\
3263-733-1 & (15.00873, 45.13101) &$1.83 \pm 0.34$ & $95.576 \pm 1.290$&$-3.277 \pm 0.425$& $14.91 \pm 1.46$ &  $5425 \pm 89$&$3.8 \pm 0.1$ &$-0.81 \pm 0.16$ & $517 \pm 55$ &$0.88 \pm 0.09$ &$12.3 \pm 2.2 $ & 0 \\
3285-1422-1 & (32.53176, 47.41257) &$1.10 \pm 0.29$ &$75.04 \pm 1.682$ & $-31.531 \pm 0.505$& $25.43 \pm 1.54$& $5214 \pm 89$ &$4.1 \pm 0.1$ & $-1.58 \pm 0.16$& $143 \pm 87$ & $0.64 \pm 0.08$& $10.9 \pm 1.3$ & 2 \\
3330-120-1 & (56.71171, 48.53692) &$2.61 \pm 0.30$ & $194.055 \pm 0.323$&$-123.109 \pm 0.255$& $-24.12 \pm 1.26$ &  $5735 \pm 89$&$3.8 \pm 0.1$ & $-1.55 \pm 0.16$& $571 \pm 30$ &$0.83 \pm 0.03$ & $12.5 \pm 0.9 $ & 0\\
3661-974-1 & (4.55758, 57.6662) &$3.49 \pm 0.651$ &$180.078 \pm 1.110$ & $104.039 \pm 0.651$& $-154.53 \pm 2.02$ &  $6507 \pm 100$& $4.1 \pm 0.2$ &$-0.99 \pm 0.16$& $397 \pm 83$ &$0.87 \pm 0.09$ & $10.2 \pm 2.7 $ & 1\\
3744-1546-1 & (67.80849, 58.96855) & $1.81 \pm 0.42$&$143.706 \pm 1.923$ &$-38.217 \pm 1.272$&  $8.72 \pm 1.49$& $6232 \pm 174$ &$4.3 \pm 0.3$ &$-1.68 \pm 0.20$ & $294 \pm 78$ &$0.78 \pm 0.05$ & $9.9 \pm 4.0 $ & 2\\
3945-1023-1 & (304.24414, 56.57186) &$-6.07 \pm 0.89$ &$-6.097 \pm 1.826$ &$-1.265 \pm 1.879$& $-18.79 \pm 1.80$& $6239 \pm 83$ & $3.8 \pm 0.2$&$-0.02 \pm 0.12$ & $1185 \pm 150$ & $1.54 \pm 0.11$& $2.3 \pm 0.5 $ & 0\\
3983-1873-1 & (338.34366, 52.68866) & $1.84 \pm 0.23$& $133.342 \pm 0.094$&$72.34 \pm 0.082$&  $-165.28 \pm 0.86$&  $4832 \pm 68$& $2.0 \pm 0.2$& $-1.27 \pm 0.14$& $1096 \pm 151$ &$1.06 \pm 0.19$ &$5.4 \pm 2.5 $ & 0\\
4032-1542-1 & (26.42901, 60.39286) &$0.74 \pm 0.40$ &$68.109 \pm 0.761$ &$-13.725 \pm 0.73$ & $-115.48 \pm 7.15$ & $7600 \pm 83$ & $3.7 \pm 0.2$&$-0.23 \pm 0.12$& $1009 \pm 187$ & $2.02 \pm 0.16$ &$0.9 \pm 0.2 $ & 0\\
4307-1106-1 & (8.16184, 74.08742) & $2.31 \pm 0.52$&$72.556 \pm 1.141$ &$15.474 \pm 1.291$& $45.88 \pm 1.79$ & $5517 \pm 74$ &$3.5 \pm 0.2$ & $-0.45 \pm 0.11$& $844 \pm 193$ & $1.41 \pm 0.20$&$3.1 \pm 2.4 $ & 0\\
4507-1461-1 & (33.29978, 82.01739) & $2.52 \pm 0.31$& $85.192 \pm 0.661$&$0.366 \pm 0.836$& $-384.65 \pm 2.22$ & $6516 \pm 100$ & $4.2 \pm 0.2$&$-1.24 \pm 0.16$ & $331 \pm 30$ &$0.82 \pm 0.02$ &$11.8 \pm 1.6 $ & 0\\
4509-1013-1 & (58.91556, 75.28116) &$2.15 \pm 0.24$ &$97.297 \pm 0.886$ &$-29.216 \pm 0.758$&  $-155.52 \pm 1.55$& $5890 \pm 89$ &$3.8 \pm 0.1$ & $-1.71 \pm 0.16$& $549 \pm 69$ & $0.83 \pm 0.08$&$ 12.0 \pm 1.9 $ & 0\\
4515-1197-1 & (79.71826, 77.83392) & $1.28 \pm 0.28$&$96.148 \pm 0.892$ &$45.051 \pm 1.045$& $-198.41 \pm 1.09$ & $5398 \pm 63$ &$3.4 \pm 0.2$ & $-1.63 \pm 0.17$& $902 \pm 170$ &$0.88\pm 0.15$ &$ 11.4 \pm 3.5 $ & 0 \\
4521-322-1 & (55.43942, 81.069) &$3.22 \pm 0.35$ & $160.469 \pm 0.536$&$1.117 \pm 0.768$& $-129.92 \pm 1.19$ & $5872 \pm 89$ & $4.0 \pm 0.1$&$-1.38 \pm 0.16$& $428 \pm 29$  & $0.83 \pm 0.02$& $ 12.4 \pm 0.6 $ & 0\\
\end{tabular}
\begin{tablenotes}

\item $^8$ This star has a very low $\log g$, making the position of the isochrones uncertain. Furthermore, its metallicity is outside of the range of our isochrones, therefore distance, mass, and age could be biased or offset.

\item \textbf{Notes:} Hipparcos and {\it Gaia} identifiers for these stars are given in Table \ref{TAB:ID} in Appendix \ref{app:GaiaID}. Proper motions and parallaxes are from {\it Gaia} TGAS, while stellar parameters have been derived using the RAVE pipeline. The 2.5 \kms uncertainty floor is \emph{not} included in the quoted HRV errors, see discussion in \S \ref{sec:stel_params}. $F$ = flag for the stellar parameter pipeline: 0 = converged; 1 = not converged; 2 = the pipeline oscillated between two solutions and the mean has been performed; 3 = bookkeping flag, the pipeline has converged.

\end{tablenotes}

\end{threeparttable}
\end{table}
\end{landscape}

where $\rho_{i,j}$ is the correlation between the parameters $i$ and $j$, as given in TGAS. We model the prior probability on distances following the ``Milky Way prior" approach presented in \cite{astraatmadjaI}. We consider a three-dimensional density model for our Galaxy, that takes into account selection effects of the {\it Gaia} survey:
\begin{equation}
\label{eq:MWprior}
P_\mathrm{MW}(d,l,b) = d^2 \rho_\mathrm{MW}(d,l,b) \ p_\mathrm{obs}(d,l,b) .
\end{equation}

The stellar number density of the Milky Way $\rho_\mathrm{MW}(d,l,b)$ is modelled as the sum of three components (see Appendix A in \cite{astraatmadjaI} for details), while $p_\mathrm{obs}(d,l,b)$ describes the fraction of observable stars in a given sky position (Equation (4) in \cite{astraatmadjaI}). We choose this prior in our analysis because it gives the best results when comparing distances with a sample of known Cepheids \citep{astraatmadjaII}. The impact of assuming different priors on distance is discussed in Appendix \ref{app:priors}: except at distances $>800$ pc, where errors are large, different priors give similar results. We assume uniform priors on proper motions. By means of Bayes' theorem we draw random samples of proper motions and distances from the resulting posterior distribution with an affine invariant ensemble Markov Chain Monte Carlo (MCMC) sampler \citep{Goodman10}, using the \emph{emcee} implementation \citep{emcee}. We run the chain with $32$ walkers and $4000$ steps per walker, for a total of $128000$ points drawn from the resulting posterior probability distribution. We check the convergence of the chain in terms of both the mean acceptance fraction and the auto-correlation time.

\begin{figure}
\centering
\includegraphics[width=0.5\textwidth]{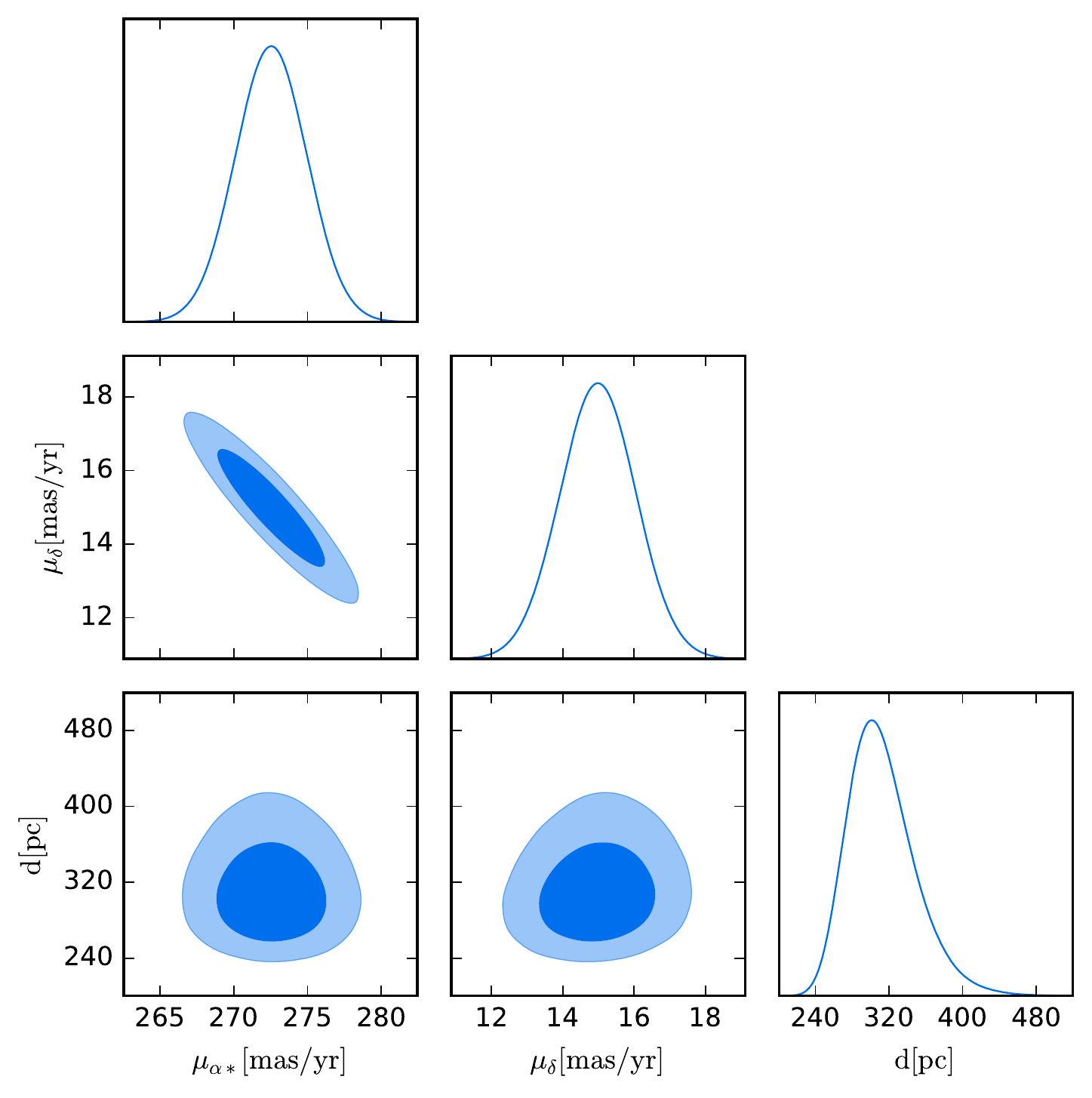}
\caption{Proper motions and distance posterior distributions for the candidate TYC 49-1326-1 as obtained from the MCMC. Correlations from TGAS are $\rho_{\mu_{\alpha *},\mu_\delta} = -0.909$, $\rho_{\mu_{\alpha *},\varpi} = 0.023$, $\rho_{\mu_\delta,\mu_\varpi} = -0.103$. Dark (light) blue regions indicate the extent of the 1$\sigma$ (2$\sigma$) credible intervals.}
\label{fig:cornerplot}
\end{figure}

An example of a cornerplot showing Bayesian posterior distributions and correlations between the astrometric parameters for the candidate TYC 49-1326-1 is shown in Figure \ref{fig:cornerplot}.

\begin{figure}
\centering
\includegraphics[width=0.5\textwidth]{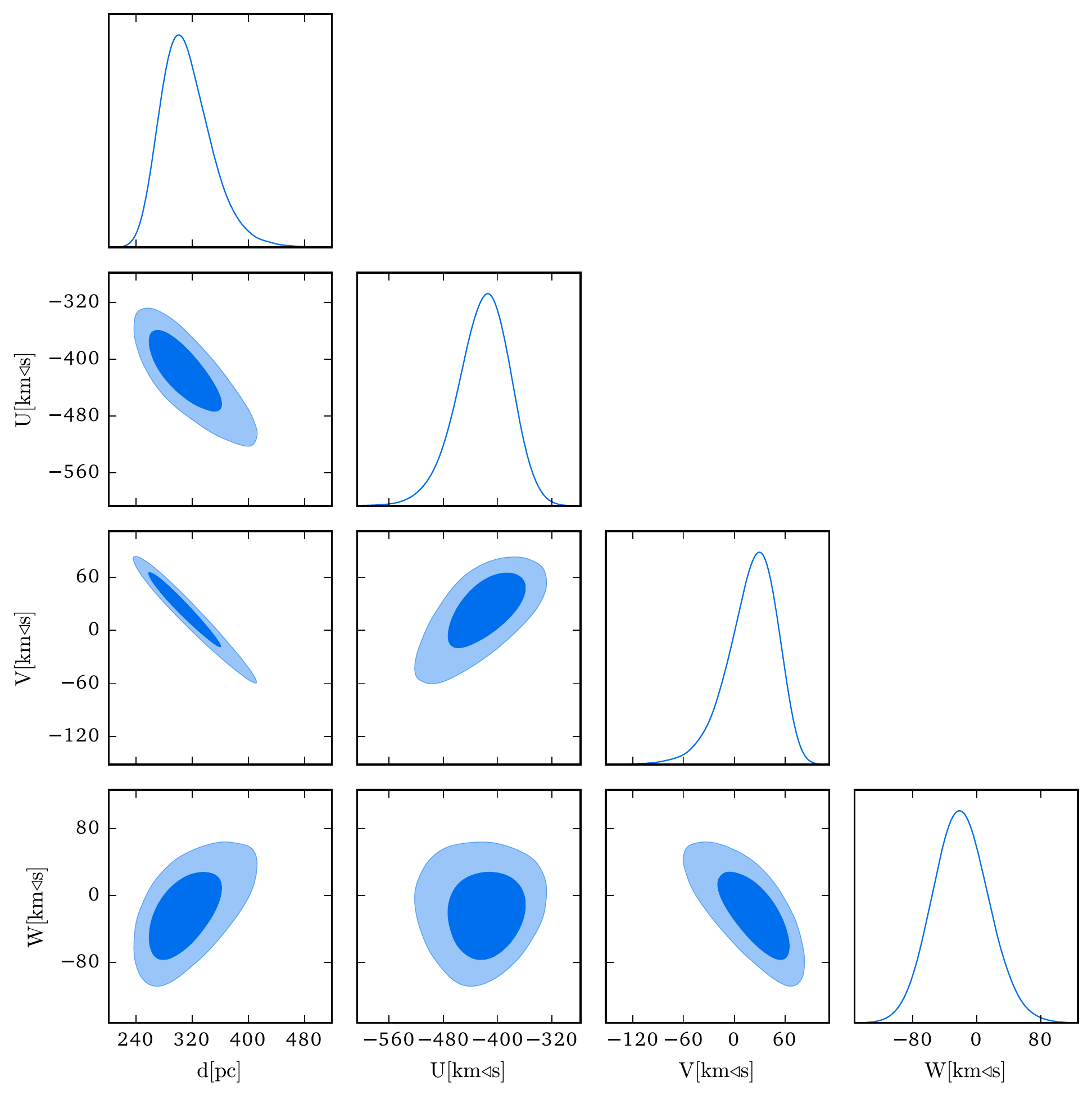}
\caption{Distance and Galactic rectangular velocities $U, V, W$ posterior distributions for TYC 49-1326-1 as obtained from the sampling of the astrometry shown in Figure \ref{fig:cornerplot}. Dark (light) blue regions indicate the extent of the 1$\sigma$ (2$\sigma$) credible intervals. The total galacto-centric velocity is $v_\mathrm{GC}=419_{-35}^{+38}$ \kms.}
\label{fig:cornerplot_UVW}
\end{figure}

For the subset of 22 stars with a spectroscopic distance estimate we simply draw proper motions from a bivariate Gaussian distribution using the $2\times 2$ covariance matrix provided by TGAS, and distances from a Gaussian with standard deviation equal to the estimated random uncertainty on distance. 

If parallax-inferred and spectroscopic distance estimates are consistent within the errors, we expect the difference between the two divided by combined uncertainties to be distributed as a Gaussian with mean of zero and standard deviation of one. If we compute a Kolmogorov-Smirnov test to check whether these two distributions are consistent, we find that the null hypothesis cannot be rejected at a 5\% level of significance. This is due to large uncertainties in distances, especially when adopting TGAS parallaxes. Since the two estimates can be remarkably different for individual stars, in the following we will present and discuss results assuming both distances.

\section{Results}
\label{sec:results}

Exploiting archival and new data we have assembled a total of 47 candidates with 3D position and velocity. A positive identification of a HVS requires both a radial trajectory from the Galactic Centre and a total velocity above the local escape speed. A star with the latter property but a trajectory that originates from the stellar disc will be called an {\it hyper runaway star}. Finally, {\it bound HVSs} (BHVSs) have Galactic Centre origin but velocity below the escape speed.

\subsection{Total Galactocentric velocity}
\label{sec:GCvel}

In order to identify HVSs, we compute the total velocity in the Galactic rest frame $v_{GC}$ for the 47 candidates with a reliable radial velocity measurement. We start correcting radial velocities and proper motions for solar and LSR motion, assuming a three-dimensional Sun's velocity vector and LSR velocity \citep{schonrich}. We then calculate Galactic rectangular velocities $U$, $V$, and $W$ with the following convention: $U$ is positive if pointing towards the GC, $V$ is positive along the direction of Galactic rotation, and $W$ is positive towards the North Galactic Pole \citep{johnson_soderblom}. The total velocity in the Galactic rest-frame is then simply computed summing in quadrature these three velocity components. We estimate uncertainties in the velocity vector via MC simulations, using the sampling in proper motions and distance described in \S \ref{sec:dist}. An example of posterior distributions for rectangular velocities is shown in Figure \ref{fig:cornerplot_UVW} for the candidate TYC 49-1326-1, obtained using posterior distributions shown in Figure \ref{fig:cornerplot}. 

\begin{figure*}
\centering
\includegraphics[width=\textwidth]{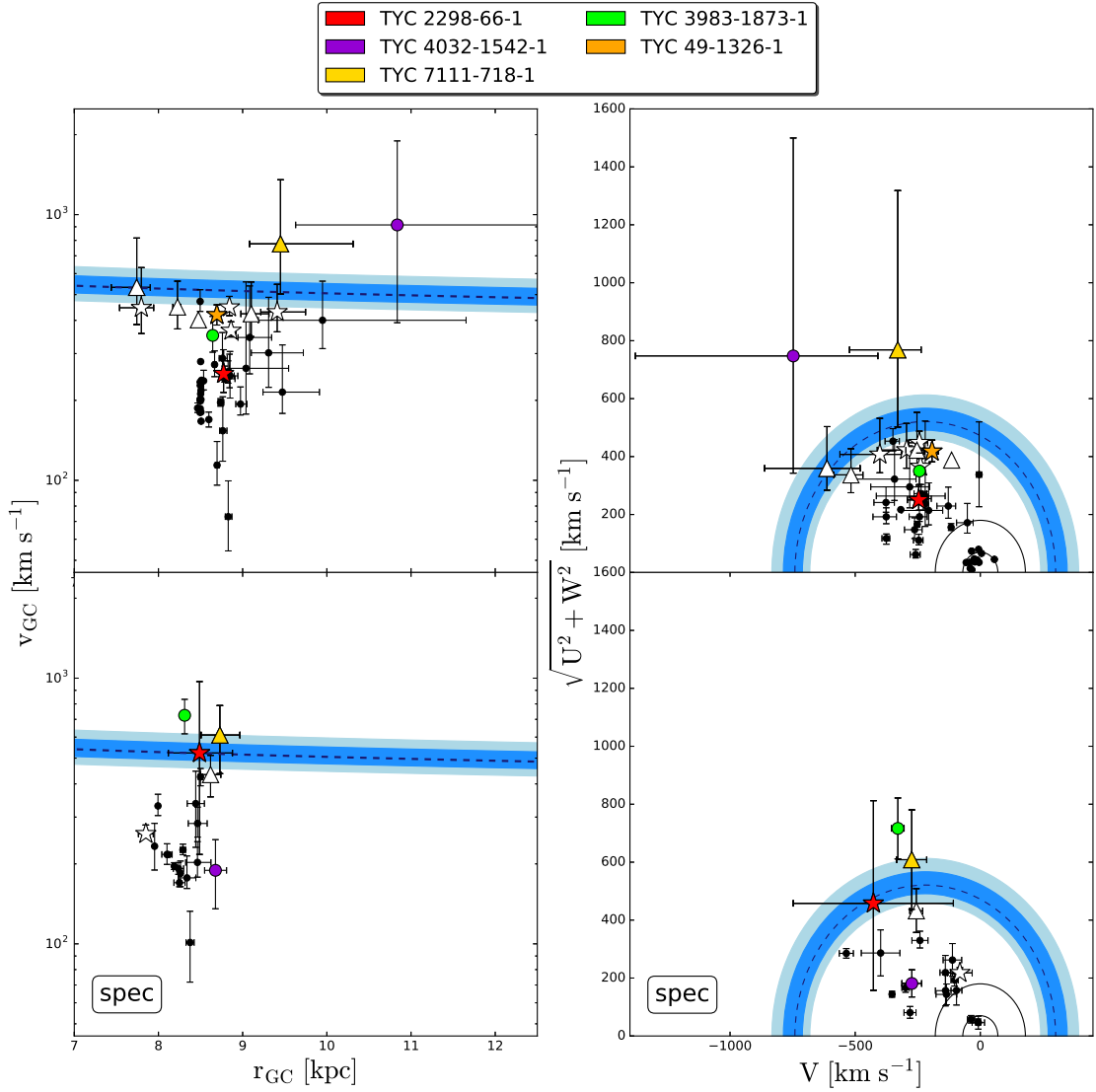}
		\caption[]{\emph{First column}: Total Galactic rest frame velocity versus Galactocentric distance for those HVS candidates with a reliable radial velocity measurement. \emph{Second column}: Toomre diagrams (in the LSR frame) for the same candidates. The two black rings in the bottom-right corner refer to the boundaries of the thin and thick disk, respectively at a constant velocity of $70$ and $180$ \kms \citep{venn+14}. Most of our candidates lie in the kinematic region corresponding to halo stars. \emph{First row}: velocities computed using distances inferred from parallax, using the MW prior. \emph{Second row}: velocities computed using a spectroscopic distance estimate, when available. \emph{All plots}:  The dashed line is the median posterior escape speed (as a function of radius in the first column, and the local $521_{-30}^{+46}$ \kms in the second one) from \cite{williams+17} with the $68\%$ ($94\%$) credible interval shown as a dark (light) blue band. Stars mark HVS/BHVS candidates in Table \ref{TAB:highV}. Triangles mark runaway star candidates in Table \ref{TAB:highV}. $11$ objects are consistent with being unbound from the Milky Way in the first row, and $3$ if we adopt spectroscopic distances.}
		\label{fig:vGC}
\end{figure*}

For each star we draw $10^5$ random realizations of its astrometric parameters, and the resulting total velocities are plotted in the first column of Figure \ref{fig:vGC} as a function of Galactocentric distance. We quote our results in terms of the median of the distribution, and errors are derived from the $16$th and $84$th percentiles. We overplot the median escape speed from the Milky Way derived in \cite{williams+17} using a dashed line, with corresponding $68\%$ ($95\%$) credible intervals shown as a dark (light) blue region. This shows how the algorithm succeeded in finding high-velocity stars: 45 out of 47 candidates have a median Galactic rest frame velocity $> 150$ \kms, which is the typical velocity dispersion of stars in the halo \citep{smith+09, evans+16}. Considering parallax-inferred distances, first row, 11 objects are compatible within their uncertainties to be unbound from the Milky Way. If we use spectroscopic estimates, we find 3 stars with a total velocity consistent with being greater than the median escape speed at their position. Discussion of individual objects is postponed to \S \ref{sec:candidates}.

Total velocities and distances are presented in Table \ref{TAB:highV} for the 15 stars with a median Galactic rest-frame velocity $> 350$ \kms obtained with at least one of the distance estimation methods. The {\it Gaia} and Hipparcos identifier of these high velocity candidates is presented in Appendix \ref{app:GaiaID}. We assign to each star its probability of being unbound from the Galaxy, $P^\mathrm{u}$. From the posterior probability on distance $d$, we can compute the escape velocity from the Galaxy in each realization of the star's position using the analytic fit in \cite{williams+17}. We define $P^\mathrm{u}$ as the fraction of Monte Carlo realizations with $v_\mathrm{GC}(d)> v_\mathrm{esc}(d)$.

In the right panels of Figure \ref{fig:vGC} we present Toomre diagrams in the LSR frame for our candidates. In a Toomre's diagram one can identify three regions (separated by two solid black lines), corresponding to stars in the thin, thick disc, and halo \citep{venn+14, hawkins+15}. In the stellar halo kinematic region we report the local escape speed with associated errors \citep[blue stripe,][]{williams+17}\footnote{We choose for simplicity to plot the local value.}. The two panels correspond to different distance determinations. Most of our candidates are consistent, from a kinematic point of view, with being halo stars. A total of 12 objects are consistent with being thin/thick disc stars considering parallax-inferred distances, and therefore will not be furthermore discussed.

\subsection{Orbital traceback}
\label{sec:traceback}

We now proceed to establish the star candidate's origin by tracing back its trajectory in different models for the Galactic potential. We decide to perform the full orbit integration only for the most promising high-velocity stars in our sample, imposing the cut $\max(v_\mathrm{GC}, \ v_\mathrm{GC spec}) > 350$ \kms, where quoted values denote the median of the distribution. A total of $15$ objects passes this cut (see Table \ref{TAB:highV}). 

We use the publicly available python package {\it galpy}\footnote{\url{http://github.com/jobovy/galpy}} \citep{galpy} to integrate the orbit of each object in the Milky Way. We run $10^5$ MC realizations of the star's orbit, using as initial conditions the position, distance, and $U$, $V$, $W$ velocities previously randomly sampled from the posterior distributions. We use a four components Galactic potential, and we study the impact of our results depending on the choice of its parameters. 

Our fiducial model consists of a point mass black hole potential: 
\begin{equation}
\label{eq:BH}
\phi_{BH}(r) = -\frac{G M_\bullet}{r},
\end{equation}
a spherically symmetric bulge modelled as a Hernquist spheroid \citep{hernquist90}:
\begin{equation}
\label{eq:bulge}
\phi_b(r) = -\frac{G M_b}{r+r_b},
\end{equation}
a Miyamoto-Nagai disc in cylindrical coordinates $(R,z)$ \citep{M&N75}:
\begin{equation}
\label{eq:disc}
\phi_d(R,z) = -\frac{G M_d}{\sqrt{R^2 + \bigl(a_d + \sqrt{z^2 + b_d^2}\bigr)^2}},
\end{equation}
and a Navarro-Frenk-White (NFW) profile for the dark matter halo \citep{nfw96}:
\begin{equation}
\label{eq:nfw}
\phi_h(r) = -\frac{G M_h}{r} \ln \Biggl(1 + \frac{r}{r_s} \Biggr).
\end{equation}
We adopt the following values for the potential parameters: $M_b = 3.4 \times 10^{10}$ M$_\odot$, $r_b = 0.7$ kpc, $M_d = 1.0 \times 10^{11}$ M$_\odot$, $a_d = 6.5$ kpc, $b_d = 0.26$ kpc \citep{johnston+95, Price2014, hawkins+15}, $M_h = 0.76 \times 10^{12}$ M$_\odot$, $r_s = 24.8$ kpc \citep{rossi+16}. This potential gives a local escape speed $\sim 580$ \kms, in agreement with results in \cite{Piffl}, and, using data presented in \cite{huang+16}, provides a good fit to the rotation curve of the Milky Way out to $\sim 100$ kpc \citep[see Appendix A, Figure A1, in][]{rossi+16}.

For those stars for which we do not have a spectroscopic estimate of the age, we trace the orbit back in time for a fiducial time of $10$ Gyr, motivated by the typical age and mass of the observed sample (see Table \ref{TAB:int} and Figure \ref{fig:Mass}). We integrate each orbit with a time resolution of $0.5$ Myr, keeping track of each disc crossing (Galactic latitude $b = 0$). 

\begin{figure}
\centering
\includegraphics[width=0.5\textwidth]{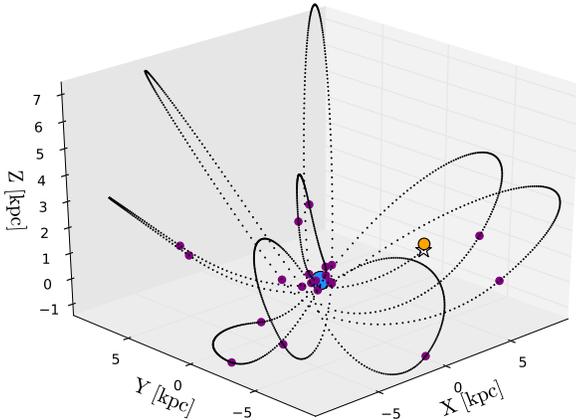}
\caption{Example MC realization of a single bound orbit of TYC 2298-66-1 using the spectroscopic distance estimate. The blue (orange) circle marks the position of the GC (Sun), and the white star corresponds to the observed position of the star. Purple dots mark the disc crossings of the star prior to, and including the one happening closest to the GC. The initial conditions are $d_0 = 1018$ pc, $v_\mathrm{GC} = 225$ \kms, the eccentricity is $e \sim 0.96$, and the estimated flight time from the assigned ejection location to the observed position is $t_\mathrm{f} = 1.3$ Gyr $\ll t_\mathrm{age} = 8.2$ Gyr. For this particular orbit, the closest disc crossing is at $\sim 260$ pc from the Galactic Centre.}  
\label{fig:orbit}
\end{figure}

If a star is ejected via the Hills mechanism but it is still gravitationally bound to the Milky Way, after the turn-around (maximum distance from the GC) it might cross multiple time the disc before being observed. This is supported by the fact that INT observations suggest that the majority of our stars have ages much larger than typical flight times from the stellar disc to the observed position, the latter being of the order of hundreds of Myr. An example of such a bound orbit is shown in Figure \ref{fig:orbit}. Thus it is not trivial to determine which disc crossing should be assigned in order to understand whether or not our candidates effectively originate from the GC. \cite{zhang+16}, searching for nearby low mass high velocity stars, assume the most-recent disc crossing to be the ejection location of the star in the Galaxy. Given the complexity of bound orbits, we simply check the consistency of the GC origin hypothesis for our candidates by recording the closest disc crossing to the GC. This approach allows us to directly exclude stars that are not HVSs, since it is a necessary condition for a HVS that this method results in a density contour level containing the GC.

We find 8 stars to have orbits consistent with coming from the Galactic Centre using parallax-inferred distances. Within the sample of stars with spectroscopic distances we find $3$ candidates, and all of them originate from the GC also when parallax-inferred distances are used.

We check the robustness of this conclusion integrating trajectories in different Milky Way potentials. Our choice for the mass of the bulge is significantly higher compared to the latest observational constraints \citep{bland-hawthorn+16, mcmillan17}, therefore we integrate each candidate assuming a bulge mass equal to half the previous adopted value: $M_b = 1.7 \times 10^{10}$ M$_\odot$, keeping fixed all the other parameters. As a second test, we adopt the potential in \cite{kenyon+14}, commonly adopted in HVS papers, which has a less massive bulge and stellar disc (but different scale parameters). In both cases we find the same candidates to be consistent with coming from the GC. As a final test, we study the impact of assuming a triaxial profile for the bulge, which might influence the orbital traceback in the inner regions of the Galaxy. Results from star counts recently revealed that the Milky Way bulge has a boxy/peanut shape \citep{mcwilliam+07, wegg+13}, which can be characterized by an axis ratio from top $(b/a) \sim 0.5$, and an edge-on axis ratio $(c/a) \sim 0.26$ \citep{bland-hawthorn+16}. Adopting the same mass and scale radius as in our fiducial potential and using a triaxial Hernquist profile to model the bulge, we find the shape of the density contour to change considerably, but the assumption of consistency with coming from the GC is solid. 

\begin{figure*}
\centering
\includegraphics[width=0.47\textwidth]{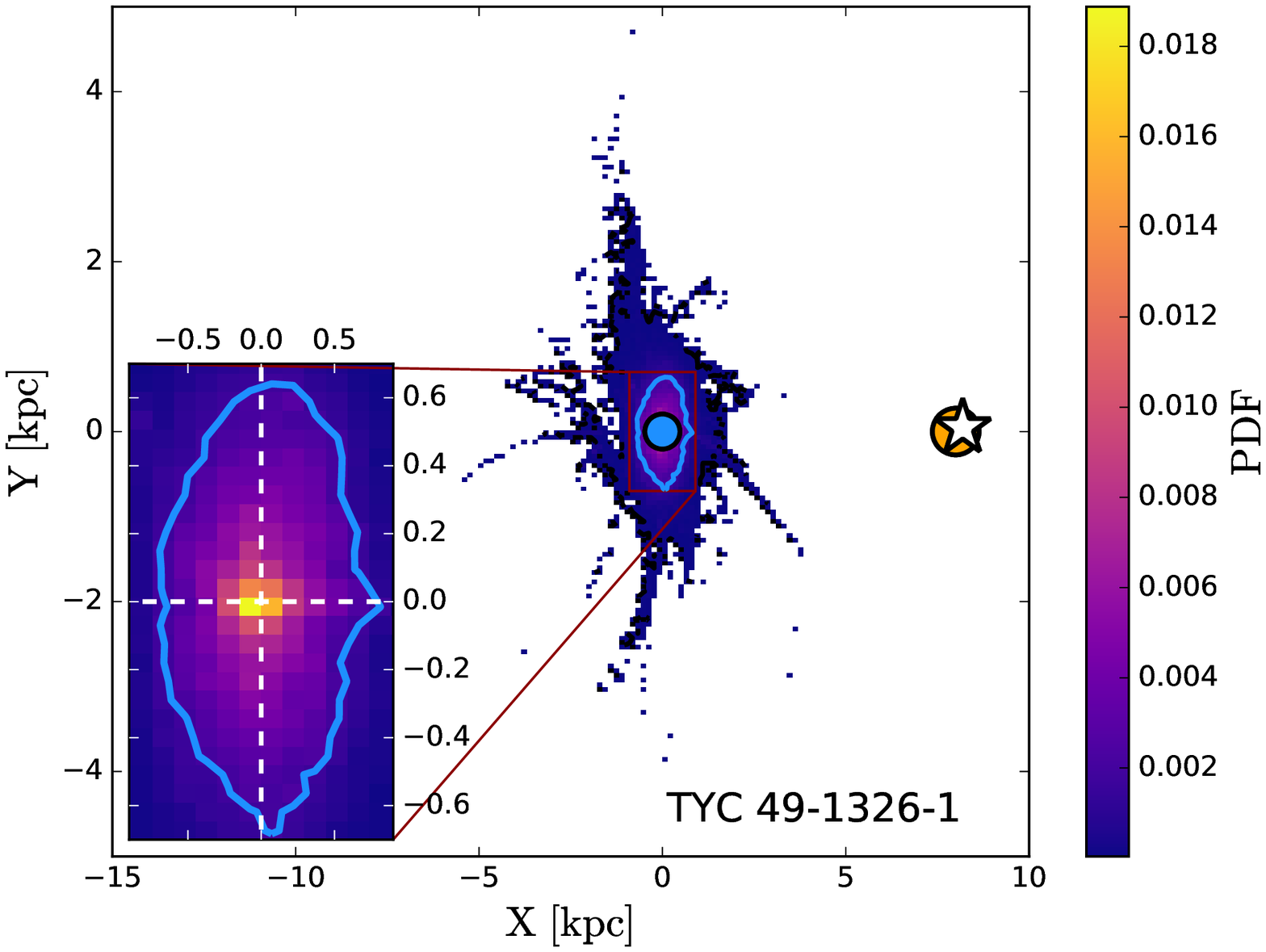} 
\qquad 
\includegraphics[width=0.47\textwidth]{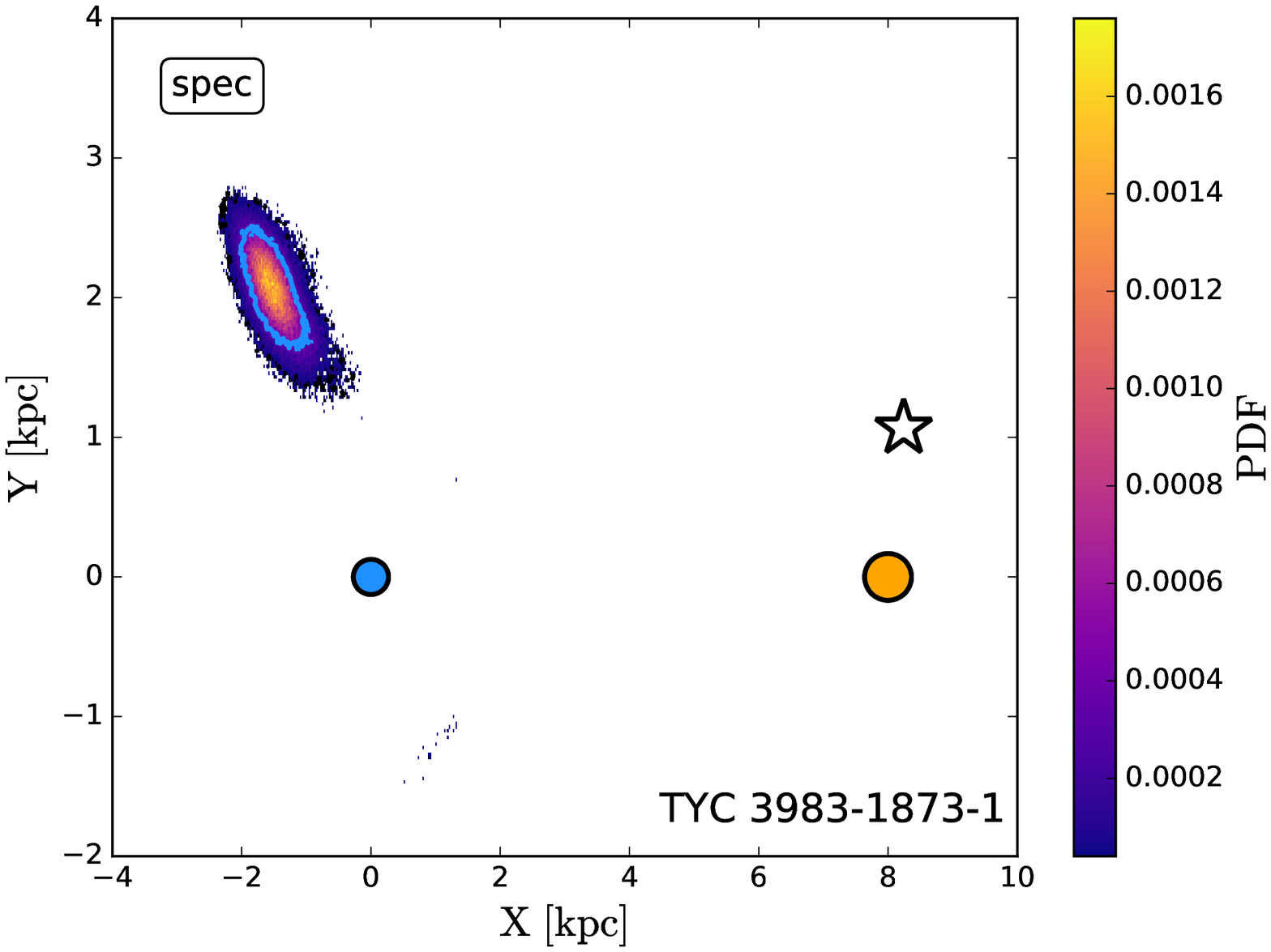}
		\caption{Normalised probability distribution function of Galactic disc crossings for the candidates TYC 49-1326-1, assuming the parallax-inferred distance (left panel), and TYC 3983-1873-1, using the spectroscopic distance (right panel). The blue line marks the $1\sigma$ contour, and the coloured region extends up to the $2\sigma$ contour. The MW rotates anticlockwise. The blue (orange) circle marks the position of the GC (Sun), while the white star corresponds to the median observed position of the candidate. The white dashed cross marks the position of the GC in the zoomed inset.
		}
		\label{fig:traceback}

\end{figure*}

Figure \ref{fig:traceback} shows example probability density functions of the disc crossing locations in the Galactic plane (rotating anticlockwise) for two candidates which will be further discussed in next sections, assuming our fiducial model for the Galactic potential. TYC 49-1326-1, left panel, is consistent with coming from the GC, while for TYC 3983-1873-1, right panel, the GC origin is excluded.

\begin{table*}
\caption{Derived kinematic properties for the 15 HVS candidates with $\max(v_\mathrm{GC}, \  v_\mathrm{GC spec}) > 350$ \kms, and interpretation.}
\label{TAB:highV}

\begin{threeparttable}

\begin{tabular}{lccccccccccr}
\hline
Tycho 2 ID & HRV & $[M/H]$ & $d$ & $d_\mathrm{spec}$ & $v_\mathrm{GC}$ &  $v_\mathrm{GC spec}$  & $P^\mathrm{u}$ & $P^\mathrm{u}_\mathrm{spec}$ & Ref \\
& (km s$^{-1}$) & (dex) & (pc) &  (pc) & (km s$^{-1}$)  & (km s$^{-1}$)  & & \\

\hline

\textbf{HVS / BHVS candidates} \\

2298-66-1 & $-31.66 \pm 2.78$ & $-2.08 \pm 0.26$ & $431_{-55}^{+78}$  &  $754 \pm 569$ & $248_{-38}^{+58}$ &  $519_{-307}^{+451}$  & $0.1$\% & $50.3$\% & 1\\[0.15cm]
8422-875-1\footnotemark   & $200.8 \pm 0.8$  & $-1.01 \pm 0.07$   & $1010_{-218}^{+400}$  &  $208 \pm 124$  & $446_{-89}^{+186}$& $259_{-7}^{+21}$  & $29.1$\% & $0.0$\% & 2, 5\\[0.15cm]
2456-2178-1 & $-243.08 \pm 49.53$ & $-2.25\pm 0.24$ & $976_{-207}^{+358}$   &                 & $430_{-68}^{+117}$ &  & $22.7$\%  &  & 3\\[0.15cm]
2348-333-1 & $205.26 \pm 0.34$ & $-1.26 \pm 0.40$   & $407_{-40}^{+51}$        &                & $448_{-32}^{+44}$  &   & $7.6$\% & & 3, 4 \\[0.15cm]
49-1326-1  & $265.1 \pm 37.6$  &  & $304_{-30}^{+38}$      &                 & $419_{-35}^{+38}$  &  & $1.2$\% & &  2, 5\\[0.15cm]
5890-971-1  & $348.6 \pm 0.8$  &   & $550_{-72}^{+93}$     &  &   $366_{-20}^{+29}$  & & $0.2$\% & &6, 7\\[0.15cm]
\hline
\textbf{Runaway star candidates} \\

7111-718-1  & $76.7 \pm 1.2$  & $-1.53 \pm 0.17$    & $1967_{-683}^{+1413}$  & $1552 \pm 430$ & $776_{-274}^{+576}$ &  $611_{-172}^{+176}$  & $82.2$\% & $70.7$\% & 2, 5\\[0.15cm]
8374-757-1  & $71.8 \pm 3.7$  &    & $832_{-179}^{+338}$    &           & $532_{-147}^{+284}$&  & $50.4$\% & & 8\\[0.15cm]
1071-404-1  & $-267.12 \pm 0.26$ & $\sim -0.5$  & $439_{-64}^{+91}$      &                 & $449_{-78}^{+113}$ & & $23.7$\% & & 4\\[0.15cm]
4515-1197-1 & $-198.41 \pm 1.09$ & $-1.63 \pm 0.17$   & $881_{-175}^{+292}$    &  $902 \pm 170$    & $423_{-76}^{+137}$&  $ 433_{-76}^{+78}$  &$23.5$\% & $15.6$\% & 1\\[0.15cm]
9404-1260-1 & $-94.9 \pm 0.6$  &   & $67.0_{-0.9}^{+1.0}$   &  &                 $402_{-4}^{+4}$    &  & $0.0$\%& & 9\\
\hline
{\textbf{Uncertain candidates}} \\
3983-1873-1  & $-165.28 \pm 0.86$  & $-1.27 \pm 0.14$  & $572_{-67}^{+88}$      &  $1096 \pm 151$   & $351_{-47}^{+64}$  &  $ 726_{-108}^{+107}$   & $1.5$\% & $97.2$\% & 1\\[0.15cm]
4032-1542-1 & $-115.48 \pm 7.15$  & $-0.23 \pm 0.12$   & $3216_{-1574}^{+2918}$ &  $1009 \pm 187$    & $918_{-527}^{+979}$&  $ 183_{-57}^{+59}$ & $75.7$\%& $0.0$\% & 1\\[0.15cm]
3945-1023-1 & $-18.79 \pm 1.80$  & $-0.02 \pm 0.12$   & $4978_{-1686}^{+2802}$ & $1185 \pm 150$   & $399_{-87}^{+162}$&  $ 215_{-4}^{+4}$  & $24.5$\% & $0.0$\% &  1\\[0.15cm]
3330-120-1 & $-24.12 \pm 1.26$ & $-1.55 \pm 0.16$ & $401_{-43}^{+56}$  &  $571 \pm 30$ & $247_{-44}^{+58}$ &  $425_{-32}^{+32}$ & $0.1$\% & $0.3$\% & 1 \\
\hline
\end{tabular}

\begin{tablenotes}

\item $^{11}$ The parallax-inferred distance $d$ is more likely to be correct for this RR Lyrae star (see Figure \ref{fig:dist_mod}), and is consistent with the value obtained using a PLZ relation (see discussion in \S \ref{sec:HVS}).

\item \textbf{Notes:} Hipparcos and {\it Gaia} identifiers for these stars are given in Table \ref{TAB:ID} in Appendix \ref{app:GaiaID}. The subscript ``$\mathrm{spec}$" refers to quantities computed using the spectroscopic distance (when available). For distances and Galactocentric velocities, results are quoted in terms of the median of the distribution with uncertainties derived from the 16th and 84th percentiles. The 2.5 \kms uncertainty floor (see discussion in \S \ref{sec:stel_params}) is \emph{not} included in the quoted HRV errors.

\item \textbf{References:} (1) This paper, observations at the INT; (2) \cite{kordopatis+13}; (3) \citep{lamostdr1}; (4) \cite{latham+02}; (5) \cite{kunder+16}; (6) \cite{przybylski78}; (7) \cite{barbier-brossat+94}; (8) \cite{kharchenko+07}; (9) \cite{holmberg+07}.

\end{tablenotes}

\end{threeparttable}

\end{table*}

\section{Discussion of Individual Candidates}
\label{sec:candidates}

\begin{figure}
\centering
\includegraphics[width=0.5\textwidth]{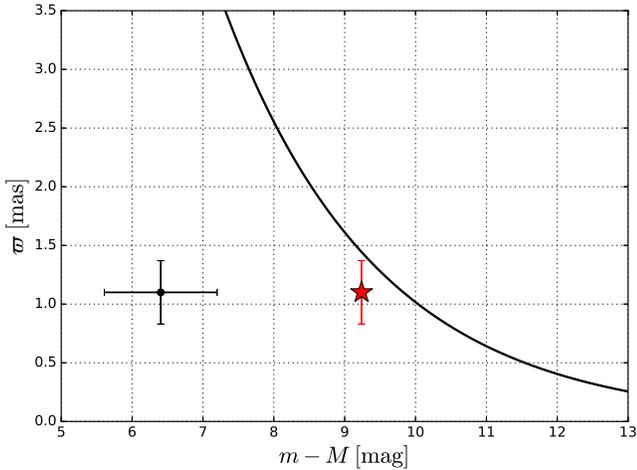}
		\caption{Parallax-distance modulus diagram for the RR Lyrae star TYC 8422-875-1 (HD 201484, V Ind), using the parallax from TGAS and the distance modulus from RAVE DR5 (black point). The line shows the analytic prediction assuming the Schlegel extinction towards the line-of-sight. The parallax-inferred distance estimate is clearly favoured. The red star corresponds to adopting the distance modulus obtained using the PLZ relation. Other candidates lie too close to the curve to have a clear preference towards one distance estimate.}
		\label{fig:dist_mod}
\end{figure} 

We divide candidates in Table \ref{TAB:highV} in three major classes: HVS and BHVS candidates, runaway star candidates, and ``uncertain" objects. To help the discussion, the metallicity distribution of these stars is shown with a purple line in Figure \ref{fig:MH}, where it is compared to typical metallicity distributions of stars in the inner Galactic halo. 
We will now discuss separately candidates from each class in detail, focusing on the most promising objects and on stars already present in literature. One additional candidate not included in Table \ref{TAB:highV}, but known from literature, is discussed in \S \ref{sec:Others}.

\subsection{HVS and BHVS Candidates}
\label{sec:HVS}

In addition to HVSs, the Hills mechanism naturally predicts a population of {\it bound HVSs}: stars having a velocity high enough to escape from the MBH's gravitational field at their ejection, but not sufficient to be unbound from the whole Milky Way. These stars, being decelerated and deflected by the Galactic potential, can cross the disc multiple times during their life, following a wide variety of highly-non-radial orbits, as previously shown in Figure \ref{fig:orbit}. The identification of such objects is observationally particularly difficult. The probability of observing a star at a particular moment of its orbit is proportional to the residence time $t_r$ in that orbit element: $p \propto t_r \propto v^{-1}$, therefore we expect most of these stars to be observed when they have low velocities, and they could thus be easily mistaken for halo stars.

Hypervelocity and bound hypervelocity star candidates are marked with a star symbol in Figure \ref{fig:vGC}. Stars are classified as HVSs if (i) their velocity is $> 350$ \kms with at least one distance estimate, and (ii) if they are consistent with coming from the GC (within $2\sigma$) when traced back in different Galactic potentials. We find a total of $6$ stars satisfying both properties within their uncertainties: TYC 2298-66-1, TYC 8422-875-1, TYC 2456-2178-1, TYC 2348-333-1, TYC 49-1326-1, and TYC 5890-971-1. The consistency with the GC origin does not depend on the assumed distance. The further sub-classification as HVSs or BHVSs depends on the value of $P^\mathrm{u}$. All of these stars are on highly radial orbits, with median eccentricities $>0.9$.

\begin{itemize}

\item TYC 2298-66-1 (LP 295-632) is a high proper motion metal-poor candidate, identified by a red symbol in Figure \ref{fig:vGC}. It is the only star with a probability $>50\%$ of being unbound from the Galaxy when using the spectroscopic distance estimate ($v \sim 530$ \kms, even if with large uncertainties), therefore it is a HVS candidate.

\item TYC 8422-875-1 (HD 201484, V Ind) is a F0 V variable star of RR Lyrae type \citep{houk78}. In the discussion of this candidate, we use Figure \ref{fig:dist_mod} to help us distinguish which distance estimate is more likely to be correct. This plot compares the position of the star in the parallax-distance modulus diagram to the analytical prediction computed assuming the Schlegel extinction towards the line-of-sight. The distance modulus is taken from RAVE DR5 \citep{kunder+16}, and the resulting point is shown in black. The total velocity of TYC 8422-875-1 strongly depends on the distance assumption, but from Figure \ref{fig:dist_mod} we can see that parallax-inferred distance is more likely to be correct. Furthermore, since this star is a RR Lyrae, we can independently determine its distance modulus using a period-luminosity-metallicity (PLZ) relation \citep{leavitt08, leavitt&pickering}. Period, $[\mathrm{Fe/H}]$ metallicity, and mid-infrared $[3.6]$ magnitude are taken from \cite{monson+17}, and we estimate the distance modulus using the PLZ relation in the {\it WISE W}1 band from \cite{sesar+17}.  This results in a distance modulus $\sim 9.3$, consistent with the parallax measured by {\it Gaia}, as shown with a red star in Figure \ref{fig:dist_mod}. We then conclude that V Ind is a BHVS candidate, with $v \sim 450$ \kms and a probability of $\sim30\%$ of being unbound.

\item TYC 2456-2178-1 is a BHVS candidate, with $v \sim 430$ \kms and a probability $\gsim 20\%$ of being unbound from the Galaxy.

\item TYC 2348-333-1 (G 95-11) is a high proper motion and high velocity star which has been previously used to estimate the local Galactic escape speed together with other stars from the $uvby - \beta$ survey of high velocity and metal poor stars \citep{garciacole+99}. With a total velocity around $450$ \kms, this star is most likely a BHVS. We note that our distance estimate is higher than the value $\sim 250$ pc given in \cite{garciacole+99}, resulting in a higher total velocity.

\item TYC 49-1326-1 (G 75-29), marked with an orange star in Figure \ref{fig:vGC}, is a BHVS candidate with a total velocity particularly well constrained of $419_{-35}^{+38}$ \kms. 

\item TYC 5890-971-1 (HD 27507), even if it has a total velocity lower than the other candidates, is worth mentioning because it is historically the first discovered HVS candidate. \cite{przybylski78} discussed the possibility that HD 27507 is a star escaping from our Galaxy given its high velocity, and a following proper motion redetermination confirmed this conclusion \citep{clements+80}. The authors found a total velocity $\sim 360$ \kms, in good agreement with our results, but studies in the past decades substantially increased the value of the local escape speed (see \cite{williams+17} for the latest constraints), making this star unlikely to be unbound from the Milky Way. Nevertheless, its orbit is consistent with coming from the GC, making TYC 5890-971-1 a bound HVS candidate.

\end{itemize}

\subsection{Runaway Star Candidates}
\label{sec:run}

Runaway stars (RSs) are high velocity stars ejected in many-body dynamical encounters in dense stellar systems \citep{poveda+67, portegieszwart00} or by the explosion of a supernova in a binary system \citep{blaauw61, tauris+98}. \cite{tauris15} showed how it is possible to reach Galactic rest frame velocities up to $\sim 1280$ \kms for the ejected companion star in a binary disrupted via an asymmetric supernova explosion. These extreme velocities can be achieved by low-mass G/K candidates in very compact presupernova binaries. High velocity runaway stars observed in the halo are most likely produced in the disc \citep{bromley+09, duarte12, kenyon+14}. Since most of our stars have masses slightly below the Solar value, this mechanism can possibly explain the notable velocity of our stars that do not originate from the GC. 

With this classification rule we identify as runaway candidates 5 high-velocity stars: TYC 7111-718-1, TYC 8374-757-1, TYC 1071-404-1, TYC 4515-1197-1, and TYC 9404-1260-1. Regardless of the adopted distance, these stars always have median $v_\mathrm{GC}> 350$ \kms. In particular, 2 stars have a probability $> 50\%$ of being unbound from the Milky Way, and are therefore classified as \emph{hyper runaway stars} (HRSs). Runaway star candidates are marked with a triangle symbol in Figure \ref{fig:vGC}. In the following we discuss them individually.

\begin{itemize}

\item TYC 7111-718-1, marked in yellow in Figure \ref{fig:vGC}, is a strong hyper-runaway star candidate, with a velocity $> 600$ \kms, in excess of the local escape speed regardless of the adopted distance estimate. From a chemical point of view, it is consistent with the inner Galactic halo population.

\item TYC 8374-757-1 (HD 176387, MT Tel) is a RR Lyrae variable star. It was previously discovered by \cite{przybylski67}, which discussed, despite large uncertainties in proper motions, its nature as a high velocity star. Because of large errors in distance we cannot strongly constrain its total velocity, which, with a median value $\sim 530$ \kms, is nevertheless consistent with being greater than the escape speed, making MT Tel a hyper-runaway star candidate. We repeat the same approach discussed for TYC 8422-875-1 to determine the distance of MT Tel using the PLZ relation in \cite{sesar+17} using data from \cite{monson+17}. We find a distance modulus $\sim 8.1$, consistent with the parallax from {\it Gaia}, confirming our high-velocity determination.

\item TYC 1071-404-1, TYC 4515-1197-1, and TYC 9404-1260-1 are RS candidates most likely bound to the MW, with a remarkably high total velocity $\gsim 400$ \kms.

\end{itemize}

Another intriguing origin for these stars not originating from the GC is that they come from the Large Magellanic Cloud (LMC), either as runaway stars \citep{boubert+17}, or by the extension of the Hills mechanism to a hypothetical MBH at the centre of the LMC \citep{boubert+16}. Uncertainties are at the moment too large to pinpoint their ejection location, and we dot not further expand on this possibility in this paper.

\subsection{Uncertain Candidates}
\label{sec:uncertain}

In our final sample (Table \ref{TAB:highV}) there are 4 stars with uncertain interpretation: TYC 3983-1873-1, TYC 4032-1542-1, TYC 3945-1023-1, TYC 2393-1001-1, and TYC 3330-120-1. These objects have a debated nature, with velocities and origins highly dependent on the assumed distance indicator. We classify as runaway star (halo star) candidates that are not consistent with coming from the GC, and with a total velocity $>350$ \kms ($<350$ \kms).

\begin{itemize}

\item TYC 3983-1873-1 (BD+51 3413) is a high proper motion HVS candidate (green points in Figure \ref{fig:vGC}). It is one of the few candidates with a spectroscopic distance higher than the parallax inferred one, which results in a total velocity of $\sim 725$ \kms, more than $1\sigma$ above the median escape speed. Remarkably, if we assume a spectroscopic distance, this object is not consistent with coming from the GC, and should therefore be classified as a HRS, while it is a BHVS candidate ($v \sim 350$ \kms) if we adopt the parallax-inferred distance.

\item TYC 4032-1542-1, marked in purple in Figure \ref{fig:vGC}, suffers from a particularly poor distance determination. The spectroscopic distance gives a relatively low velocity of $\sim 190$ \kms, consistent with that of a high velocity halo star. Its velocity increases considerably if we rely on the much more uncertain parallax-inferred distance ($v \sim 900$ \kms). A point worth mentioning is that the metallicity is considerably higher than the mean value in the inner halo, making this object worth inspecting in order to constrain its nature and origin as kinematic and chemical outlier. Furthermore, TYC 4032-1542-1 is an A type star, more massive compared to the other candidates, therefore it is more difficult to explain its high velocity invoking the disruption of a close binary via supernova explosions \citep[][and see discussion in \S \ref{sec:run}]{tauris15}.

\item TYC 3945-1023-1 is a RS ($v\sim400$ \kms) or a halo star ($v\sim 200$ \kms) candidate, if we assume the parallax-inferred or the spectroscopic distance estimate respectively.

\item TYC 3330-120-1 is a runaway star candidate ($v\sim425$ \kms) if we adopt the spectroscopic distance, but behaves as a typical halo star ($v\sim250$ \kms) if we infer distance from parallax.

\end{itemize}

\subsection{HD 5223: Most Likely Not a HVS}
\label{sec:Others}

In this subsection we present one additional star discovered with our data mining algorithm, TYC 1739-1500-1 (HD 5223). Even if it doesn't pass the velocity cut in Table \ref{TAB:highV}, this star was previously known and discussed for its high velocity, which we now revisit using {\it Gaia}'s much more precise data.

HD 5223 is a carbon-enhanced metal-poor star presented in \cite{pereira+12}, which concluded that this object is a hypervelocity star with a total velocity in the Galactic frame of $713$ \kms. Our velocity determination $v = 288_{-46}^{+72}$ \kms is considerably lower because of a substantial difference in the assumed distance: \cite{pereira+12} determined $d = 1.2$ kpc, while our computation seems to suggest lower values: $d = 565_{-80}^{+117}$ pc. If our estimate is correct, HD 5223 is bound to the MW, and furthermore we find its orbit not to be consistent with coming from the GC.

\section{Discussion and Conclusions}
\label{sec:discussion}

We successfully developed a new automatized method to extract high velocity stars, using a data-driven algorithm trained on mock populations of hypervelocity stars. Our data mining routine, an artificial neural network, is optimized for the very unbalanced search of rare objects in a large dataset. This approach avoids a bias towards particular spectral types or stellar properties, making as few assumptions as possible on the stellar nature of stars coming from the Galactic Centre.
Applying the algorithm to the TGAS subset of the first release of the {\it Gaia} satellite, we have identified a total of $80$ objects with a predicted probability $> 90\%$ of being a HVS, and for 30 of those we were able to find a radial velocity measurement from literature. We followed up spectroscopically $22$ candidates at the Isaac Newton Telescope, for a total of 47 stars with a reliable radial velocity determination. Our stars show a uniform distribution across the sky, showing that the algorithm is not selecting sources in a preferential direction.

With a Bayesian approach we inferred distances from parallax for all our candidates, and total velocities in the Galactic rest frame were computed in order to establish their nature and origin. Without pre-selection of data we were able to recover several objects already noted and discussed in literature because of their remarkably high velocities. We found $45$ candidates with a median rest frame velocity $> 150$ \kms, $14$ of them having $v > 400$ \kms, and a subset of $5$ stars has a probability $>50\%$ of being unbound from the Milky Way, with median velocities up to $\sim 900$ \kms. 

Tracing back orbits with Monte Carlo simulations in different Galactic potentials we found:
\begin{itemize}

\item $6$ stars being consistent with coming from the Galactic Center. One of these stars, with a velocity of $\sim 520$ \kms, has a probability $> 50\%$ of being unbound from the Galaxy (HVS), while the others are bound hypervelocity star candidates, with velocities $>360$ \kms;

\item $5$ stars with high velocities but trajectories not consistent with coming from the Galactic Centre: these stars are runaway star candidates. Two of these stars have probabilities $> 50\%$ of being unbound from the Milky Way, and are therefore classified as hyper runaway stars. The explosion of a supernova in a binary system is a plausible mechanism for having accelerated these stars to such high velocities. It is remarkable that a good fraction of our RS candidates have velocities consistent with being higher than the escape velocity from the Galaxy, since these stars are thought to be extremely rare: approximately $1$ for every $100$ HVSs \citep{bromley+09, perets12, kenyon+14, brown15};

\item $4$ stars with a velocity and origin highly dependent on the assumed distance estimate. Two of these stars have a high probability of being unbound from the Milky Way.

\end{itemize}

At the moment, positive identifications are strongly hampered by large uncertainties in distance and limited information on the age and flight time of our sources. The advent of future {\it Gaia} releases will dramatically increase the number of HVSs we expect to find. The more accurate parallax determination, less affected by systematics, will allow us to decrease error bars and to identify in a clearer way the most interesting objects, narrowing down their ejection location. The brightest stars in the catalogue will also have a radial velocity measurement, allowing us to train the neural network adding this precious information as an extra feature to the astrometric solution.

We are currently working to increase the quality of the training set of mock HVSs, considering not only radial trajectories, but modelling orbits of bound stars and including deviations due to the disc and to a possible triaxiality of the bulge \citep[e.g.][]{mcwilliam+07} and/or the halo \citep[e.g.][]{bullock02, helmi04}. Another natural advancement would be to model runaway and halo stars to create mock populations, and then to perform a multiclass classification analysis in order to decrease the number of false positives and achieve a more precise classifier.

\section*{Acknowledgements}

We thank J. Brinchmann and A. Patruno for useful discussion and comments, U. Bastian and L. Lindegren for suggestions and advice on the use of {\it Gaia} astrometric data and on distances determination, and T. Astraatmadja and C. Bailer-Jones for the implementation of the Milky Way Prior. We also thank Warren Brown for the careful reading of the manuscript and his useful comments. TM and EMR acknowledge support from NWO TOP grant Module 2, project number 614.001.401. ES and KY gratefully acknowledge funding by the Emmy Noether program from the Deutsche Forschungsgemeinschaft (DFG). This work has made use of data from the European Space Agency (ESA) mission {\it Gaia} (\url{http://www.cosmos.esa.int/gaia}), processed by the {\it Gaia} Data Processing and Analysis Consortium (DPAC,
\url{http://www.cosmos.esa.int/web/gaia/dpac/consortium}). Funding for the DPAC has been provided by national institutions, in particular the institutions participating in the {\it Gaia} Multilateral Agreement. The Isaac Newton Telescope is operated on the island of La Palma by the Isaac Newton Group of Telescopes in the Spanish Observatorio del Roque de los Muchachos of the Instituto de Astrof\'{\i}sica de Canarias. This research made use of Astropy, a community-developed core Python package for Astronomy \citep{astropy}. All figures in the paper were produced using matplotlib \citep{matplotlib}.

\bibliographystyle{mnras}
\bibliography{hvs_dr1}

\appendix

\section{{\it Gaia} Identifiers}
\label{app:GaiaID}

In Table \ref{TAB:ID} we present Tycho 2, Hipparcos, and {\it Gaia} identifiers for the candidates observed at the INT (Table \ref{TAB:int}) and for the stars with $v>350$ \kms (Table \ref{TAB:highV}).

\begin{table}
\caption{Tycho 2, Hipparcos, and {\it Gaia} identifiers of stars observed at the INT and of high velocity candidates.}
\label{TAB:ID}
\begin{tabular}{lcl}
\hline
Tycho 2 ID & Hipparcos ID & {\it Gaia} ID  \\

\hline

1071-404-1  & 98492  & 4299974437593772672\\
2282-208-1  &        & 314799593600582656 \\
2292-1267-1 &        & 316401685121779712 \\
2298-66-1   &        & 317585859144818688\\
2320-470-1  &        & 329685915888890880\\
2348-333-1  &        & 137859551029399040\\
2376-691-1  &        & 172747742173867904\\
2393-1001-1 &        & 180650104040989568\\
2456-2178-1 &        & 893048667206860800\\
2818-556-1  &        & 347908809291960832\\
2822-1194-1 &        & 348293878879518848\\
3163-1181-1 &        & 2081319505008076416\\
3263-733-1  &        & 377741720849393920\\
3285-1422-1 &        & 353451584846863104\\
3330-120-1  & 17648  & 248695099116287872\\
3661-974-1  &        & 422054582068454016\\
3744-1546-1 &        & 470781741956237696\\
3945-1023-1 &        & 2187713404073484288\\
3983-1873-1 & 111334 & 2000722382112691456\\
4032-1542-1 &        & 509654254003883776\\
4307-1106-1 &        & 539315160710386944\\
4507-1461-1 &        & 569097391651702656\\
4509-1013-1 &        & 550795677011227648\\
4515-1197-1 &        & 552553933541803008\\
4521-322-1  &        & 568189573004745472\\
49-1326-1   &        & 2503868695508755840\\
5890-971-1  & 20214  & 3172032703298013696\\
7111-718-1  &        & 5590900663125136000\\
8374-757-1  & 93476  & 6662886601414152448\\
8422-875-1  & 104613 & 6483680327939151488\\
9404-1260-1 & 46120  & 5195968559017084160\\
\hline
\end{tabular}

\end{table}

\section{Assuming Different Priors on Distance}
\label{app:priors}

One could argue that assuming a three-components stellar density (bulge + disc + halo) for our Galaxy $\rho_\mathrm{MW}(\bm{d})$, as in Equation \ref{eq:MWprior}, is not appropriate to model the spatial distribution of HVSs, a population of stars that, by definition, is not distributed according to the density profile of the Milky Way. Therefore in this appendix we discuss the implication of assuming different priors on distances $P(\bm{d})$ in the MCMC sampling described in \S \ref{sec:dist}. In practice we adopt two different priors and we test the impact of these choices on our results: an exponential decreasing prior $P_\mathrm{exp}(d)$, and a prior specifically tailored for HVSs, the \emph{HVS prior} $P_\mathrm{HVS}(\bm{d})$, that we introduce in this paper.

\begin{figure}
\centering
\includegraphics[width=0.5\textwidth]{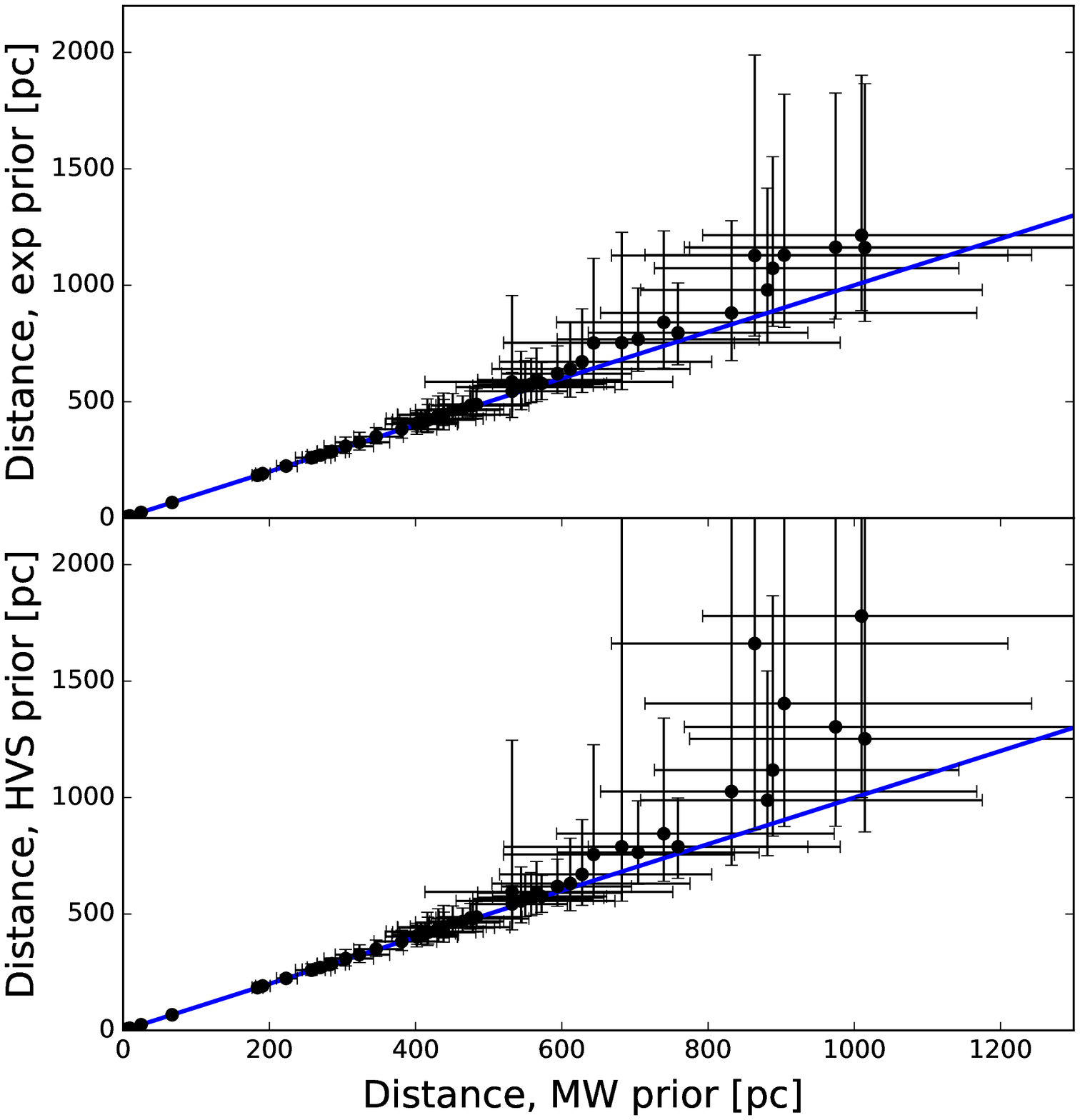}
		\caption{Comparison of distances obtained using the MW prior, on the $x$-axis, and the exponential decreasing (HVS) prior, $y$-axis on the top (bottom) panel. The blue line corresponds to equal estimates.
		}
\label{fig:dist_comparison}
\end{figure}

\cite{astraatmadjaI} show that an exponential decreasing prior
\begin{equation}
\label{eq:expprior}
P_\mathrm{exp}(d) \propto d^2 \exp\Bigl(-\frac{d}{L}\Bigr)
\end{equation}
with $L = 1.35$ kpc gives a better performance in terms of RMS errors compared to the MW prior, when resulting distance estimates are compared with GUMS simulated data. This choice assumes that the disc has the same scale-height as the scale-length, and clearly it is not an accurate description of the MW. We find that this prior overestimates distances for the majority of our candidates, with values well above the spectroscopic ones. This is evident in top panel of Figure \ref{fig:dist_comparison}, where for distances greater than $\sim 600$ pc we can see that median values obtained with the exponential prior are always higher than the ones derived with the MW prior. This is due to the choice of $L$, which sets the exponential cut-off of the distribution. Since $L=1.35$ kpc is higher than the typical distance of stars in the TGAS calatogue, this prior biases our candidates towards greater distances, and thus towards higher total velocities, proper motions and radial velocities being equal.

Assuming a continuous and isotropic ejection of HVSs from the Galactic Centre, the number density of HVSs goes approximately as $1/r^2$, where $r$ is the galactocentric radius \citep{brown15}. Following Equation \ref{eq:MWprior} we therefore construct the HVS prior as:
\begin{equation}
\label{eq:HVSprior}
P_\mathrm{HVS}(d,l,b) \propto \Biggl(\frac{d}{r(d,l,b)}\Biggr)^2  p_\mathrm{obs}(d,l,b),
\end{equation}
with $r(d,l,b) = \sqrt{d^2 + d_\odot^2 - 2dd_\odot\cos(l)\cos(b)}$ and $d_\odot=8$ kpc. When deriving distances and total velocities with this prior, we find again results to be consistent with the ones derived using the MW prior, but uncertainties are considerably larger, and this prior overestimates distances for further stars, as shown in bottom panel of Figure \ref{fig:dist_comparison}.

In the end, we choose to adopt the MW prior for presenting our results since it allows us to maintain a conservative approach: because of large uncertainties, we only interpret our candidates as HVSs at the end of the kinematic analysis, without biasing our distances and velocities using that assumption.

\label{lastpage}

\end{document}